\documentclass[12pt]{article}
\usepackage{graphicx,psfrag}
\usepackage{bm}
\begin{document}
\title{Periodic orbits, basins of attraction and chaotic beats 
in two coupled Kerr oscillators.}
\author{I. \'{S}liwa$^{a}$ and K. Grygiel $^{b}$ \\
\em{$^{a}$Theory of Nanostructures Laboratory,}\\
\em{Institute of Molecular Physics,}\\ 
\em{Polish Academy of Sciences,}\\
\em{ul. M. Smoluchowskiego 17, PL 60-179 Pozna\'{n}, Poland} \\
\em{$^{b}$Nonlinear Optics Division, Department of Physics,} \\
\em{A. Mickiewicz University} \\
\em{ul. Umultowska 85, PL 61-614 Pozna\'{n}, Poland }}
\date{\today}
\maketitle

\begin{abstract}
Kerr oscillators are model systems which have practical
applications in nonlinear optics. Optical Kerr effect i.e.
interaction of optical waves with nonlinear medium with
polarizability $\chi^{(3)}$ is the basic phenomenon needed to
explain for example the process of light transmission 
in fibers and optical couplers. In this paper we analyze the
two Kerr oscillators coupler and we show that there is a 
possibility to control the dynamics of this system, especially by
switching its dynamics from periodic to chaotic motion and vice
versa. Moreover the switching between two different stable
periodic states is investigated. The stability of the system is
described by the so-called maps of Lyapunov exponents in
parametric spaces. Comparison of basins of attractions between two
Kerr couplers and a single Kerr system is also presented.
\end{abstract}

\vspace{0.5cm}
\newpage
\section{Introduction}

One of the best known and most intensively studied optical models is an oscillator 
with Kerr nonlinearity. Different kinds of anharmonic Kerr oscillators have 
also been used to study classical and quantum chaos ~\cite{milburn}-~\cite{leo2}. 
Mutually coupled Kerr oscillators can be successfully used for a study of couplers,
 the systems consisting of a pair of coupled Kerr fibres.  The first two-mode 
Kerr coupler has been proposed by Jensen ~\cite{jensen} and investigated in 
depth in ~\cite{jensen,kenkre}. Kerr couplers affected by quantization can 
exhibit various quantum properties such as squeezing of vacuum fluctuations, 
sub-Poissonian statistics, collapses and revivals ~\cite{chefles,fiurasek}.

In the last two decades since the publication of the paper by Pecora and Carroll
~\cite{pecora} the phenomenon of synchronization in systems of
the coupled oscillators has become a subject of
comprehensive investigation. The problem of synchronization of two
linearly coupled Kerr oscillators has been studied in ~\cite{KG} and
the possibility of synchronization of chaotic motion was proved
numerically. Moreover the case of synchronization of two kinds of
Kerr couplers having a structure of low-dimensional chains (ring
and open) has been analysed ~\cite{PS}.

This paper is an attempt at using the modern
tools of nonlinear science for numerical investigation of dynamics of a system made of two coupled Kerr oscillators.

In Sec. 2 the basic equation of motion for the single Kerr
oscillator is introduced. Simple periodic solutions of
equations of motions have been found and the dynamics of the
system as well as basins of attraction for such solutions are
investigated. Moreover, we calculate the Lyapunov maps for the single 
Kerr oscillator with external periodic as well as
modulated fields. The second case of the external field is
used to generate the so-called chaotic beats. In
Sec. 3 our single Kerr system analysis is extended over the case of
two coupled Kerr sub-systems with nonlinear coupling. We find
the analytic periodic solutions of such a system. Lyapunov
maps and the basins of attraction for this system are helpful
tools in analysis of properties of these system. We find that it is
possible to change the periodic states of one sub-system by changing the initial conditions of the other sub-system (switching the periodic dynamics). Moreover, it is proved that coupled Kerr sub-systems are able to generate of chaotic beats.

\section{The single Kerr oscillator}

\subsection{Equations of motion}

We study the dynamical system described by the following
hamiltonian:
\begin{equation}\label{ker1}
H=H_{0}+H_{1}\,,
\end{equation}
where:
\begin{eqnarray}\label{ker1a}
H_{0} & = & \omega a^{*}a+\frac{1}{2}\epsilon a^{*2}a^{2}\,, \\
\label{ker1b} H_{1} & = & iF\left(a^{*}e^{-i\Omega_{p}t}
  -ae^{i\Omega_{p}t}\right)\,.
\end{eqnarray}
The hamiltonian $H_{0}$ represents the so called Kerr
oscillator (if $\epsilon =0$, then $H_{0}$ refers to the
harmonic oscillator), whereas the hamiltonian $H_{1}$ describes
the interaction of the Kerr oscillator with the periodic external
field. The quantities $a$ and $a^{*}$ are complex dynamical
variables describing the amplitudes, $\omega $ denotes the frequency of the 
free vibrations of the harmonic oscillator - basic frequency, $\epsilon $ 
is the parameter describing the Kerr nonlinearity in the system (this is the
nonlinearity of the third order) and $F$ is the external field amplitude 
at the frequency $\Omega_{p}$.

The equation of motion for variable $a$ has the form:
\begin{equation}\label{ker2}
\frac{da}{dt}= -i\omega a-i\epsilon
a^{*}a^{2}+Fe^{-i\Omega_{p}t}-\gamma a\,.
\end{equation}
The term $-\gamma a$ - added on phenomenological grounds - describes the
mechanism of loss with the damping constant $\gamma $. All the
parameters, that is $\omega $, $\epsilon $, $F$, $\Omega_{p}$ and $\gamma $ 
are taken to be real. The equation of motion for $a^{*}$ is simply a 
complex conjugation of Eq.(\ref{ker2}).

In the autonomous and conservative case, that is when $\gamma=F=0$, the 
solution of equation (\ref{ker2}) has a well-known
form:
\begin{equation}\label{ker3}
a(t)=a_{0}e^{-i(\omega +\epsilon a_{0}^{*}a_{0})t}\,,
\end{equation}
where $a(t)\mid_{t=0}=a_{0}$ is the initial condition.

In the nonautonomous case of Eq.(\ref{ker2}) we can find the periodic solution:
\begin{equation}\label{naut}
a(t)=xe^{-i(\omega +\epsilon x^{*}x)t}\,,
\end{equation}
that is in the form of the solution of the autonomous one. 
Function (\ref{naut}) satisfies the equation of motion
(\ref{ker2}) provided that $x=F/\gamma $. As a result the periodic
solution of equation (\ref{ker2}) has the form:
\begin{equation}\label{rozw1}
a(t)=\frac{F}{\gamma }e^{-i\left(\omega +\epsilon
\frac{F^{2}}{\gamma^{2}}\right)t}\,.
\end{equation}
Formally, the function in the form of (\ref{rozw1}) is the solution of the
differential equation (\ref{ker2}) only if two condition are fulfilled: (A)
$\Omega_{p}=\omega +\epsilon \frac{F^{2}}{\gamma^{2}}$, and (B)
the initial condition has the form $a(0)\equiv a_{0}=F/\gamma $. In other words, 
the periodic solution (\ref{rozw1}) is correct only for special choice of 
the set of parameters $\omega, \Omega_{p}, F, \gamma$. Finally, it is worth 
noting that in the nonautonomous case the period of solution (\ref{rozw1}) 
depends on the initial condition, as well as in the autonomous case. 
In the phase plane $(\mathit{Re}\,a,\,\mathit{Im}\,a)$ the periodic solution
(\ref{rozw1}) satisfies the phase equation (circle):
\begin{equation}\label{circle}
(\mathit{Re}\,a)^{2}+(\mathit{Im}\,a)^{2}=F^{2}/\gamma^{2}\,.
\end{equation}
for any values of frequency $\omega$.

It should be emphasized that the method presented here is usefull to find only the one periodic solution for a given set of the system parameters. Generally, the Eq. (4) have up to three (not only periodic) solutions.

\subsection{The dynamics of the system in the phase space}

\begin{figure}
\begin{center}
\includegraphics[width=7cm,height=7cm,angle=0]{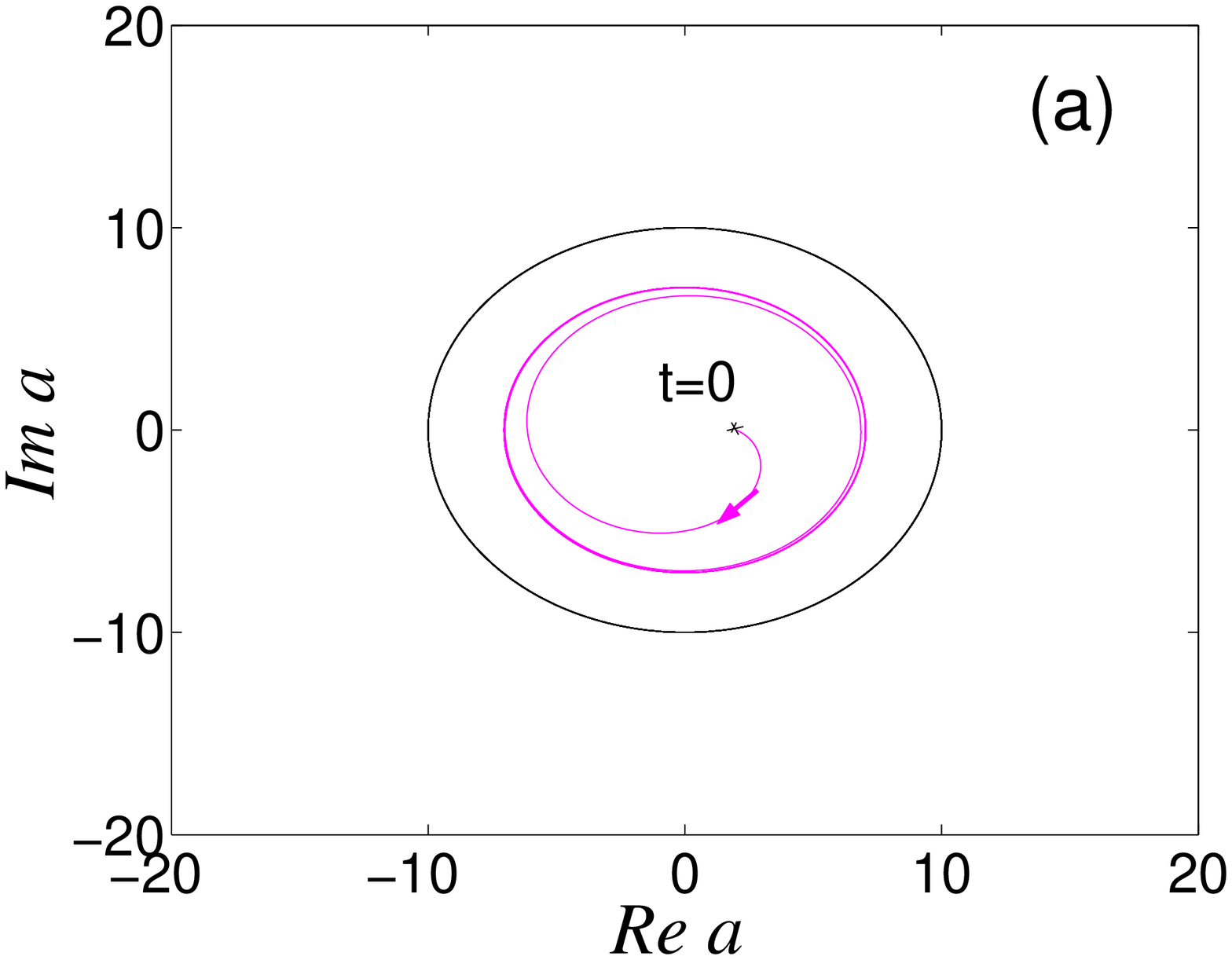}
\includegraphics[width=7cm,height=7cm,angle=0]{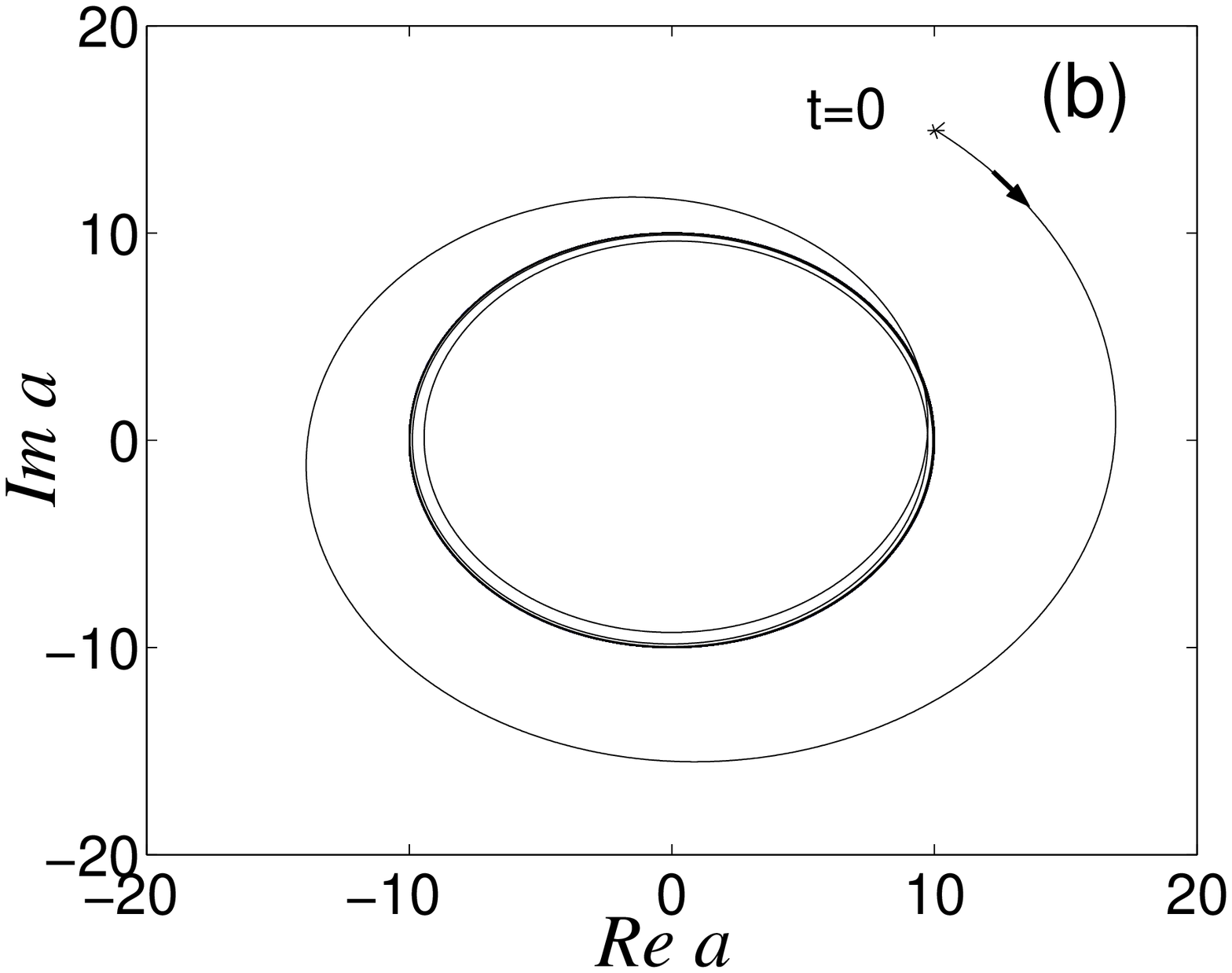}
\caption{The phase trajectories of the system (\ref{ker2}) for
$\omega =1$, $\gamma =0.5$, $\epsilon =0.01$, $F=5$ and
$\Omega_{p}=2$ and for the initial conditions: (a)
$\mathit{Re}\,a=2$, $\mathit{Im}\,a=0$; (b) $\mathit{Re}\,a=10$,
$\mathit{Im}\,a=15$.}
\label{fig1}
\end{center}
\end{figure}

As a numerical example let us consider the dynamics of a system
described by equation (\ref{ker2}), if $\omega =1$, $\gamma =0.5$,
$\epsilon =0.01$, $F=5$ and $\Omega_{p}=2$. 
Then, in compliance with (\ref{rozw1}), the periodic solution of
equation (\ref{ker2}) has the form:
\begin{equation}\label{ust2}
a(t)=10e^{-2it}\,,
\end{equation}
and in the phase space it satisfies the following equation:
\begin{equation}\label{ust3}
(\mathit{Re}\,a)^{2}+(\mathit{Im}\,a)^{2}=100\,.
\end{equation}
As a result, for the initial condition $a_{0}=10$ the phase point
draws simply a circle described by equation (\ref{ust3}). But if
the system (\ref{ker2}) starts from another initial conditions, we
observe the following interesting behaviour: after some time the
phase point tends to one of the two orbits:
$(\mathit{Re}\,a)^{2}+(\mathit{Im}\,a)^{2}=100$ or
$(\mathit{Re}\,a)^{2}+(\mathit{Im}\,a)^{2}=50$. Two examples
of such behaviour of the phase point are illustrated in Fig.
\ref{fig1}.
In Fig. \ref{fig1}(a) the phase point (marked in pink) starts from
the initial condition $\mathit{Re}\,a=2$, $\mathit{Im}\,a=0$ and
after the time $t=50$ goes into the orbit of the radius
$r=\sqrt{50}$. However in Fig. \ref{fig1}(b) we can see how the
phase point starting from the initial condition $\mathit{Re}\,a=10$,
$\mathit{Im}\,a=15$ goes into the orbit of the radius $r=10$,
described by equation (\ref{ust3}). It should be emphasized that the orbit 
$r=\sqrt{50}$ is also periodic but it does not fulfill conditions (A) and (B).

Both orbits: $(\mathit{Re}\,a)^{2}+(\mathit{Im}\,a)^{2}=100$ and
$(\mathit{Re}\,a)^{2}+(\mathit{Im}\,a)^{2}=50$ are attractors of
the system (\ref{ker2}). It means that for the 
parameters: $\omega =1$, $\gamma =0.5$, $\epsilon =0.01$, $F=5$
and $\Omega_{p}=2$ the system (\ref{ker2}) tends to one of the
two steady states (periodic), represented by these two orbits.

It is easy to show that the orbit described by equation
$(\mathit{Re}\,a)^{2}+(\mathit{Im}\,a)^{2}=50$ is identical with
that generated by equation (\ref{ker2}) if $\omega =1$, $\gamma =0.5$, 
$\epsilon =0.01$, $F=5$ and
$\Omega_{p}=1$ and for the initial condition: $\mathit{Re}\,a=5$,
$\mathit{Im}\,a=5$. This solution has the form:
$a(t)=(5+5i)e^{-2it}$. 

\subsection{Basin of attraction}

\begin{figure}
\begin{center}
\includegraphics[width=7cm,height=8cm,angle=270]{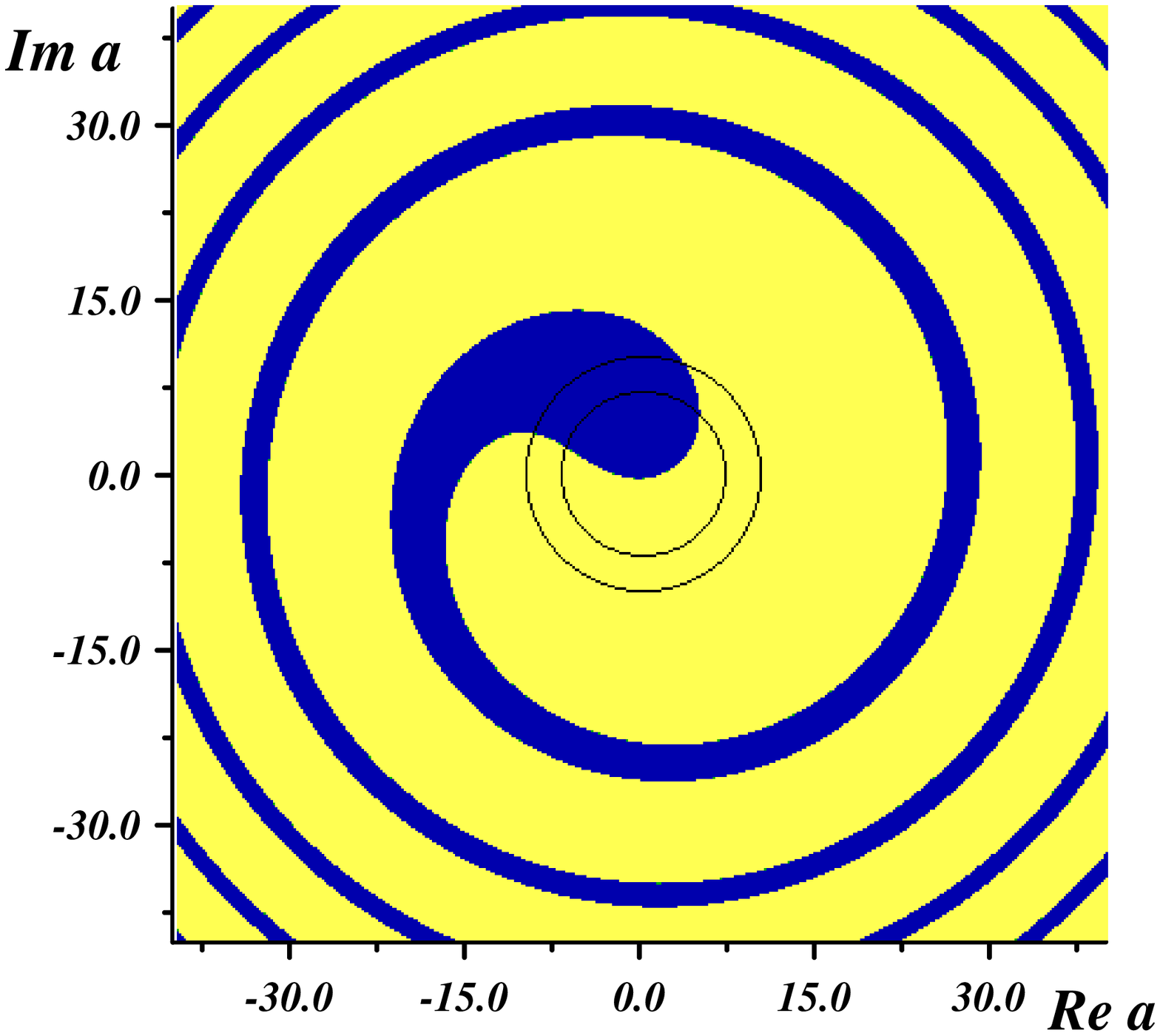}
\caption{Basins of attraction for two attractors - limit cycles
for $r=10$ and $r\prime =\sqrt{50}$. The parameters of the system
(\ref{rozw1}) are: $\omega =1$, $\gamma =0.5$, $\epsilon =0.01$,
$F=5$ and $\Omega_{p}=2$.} \label{basins1}
\end{center}
\end{figure}

To illustrate the full influence of the initial
conditions on the evolution of the system we used the so called
{\em basins of attraction}. Basin of attraction is the set of
initial conditions which lead to the system's attractor. The
basins of attraction of the system (\ref{ker2}) for $\omega =1$,
$\gamma =0.5$, $\epsilon =0.01$, $F=5$ and $\Omega_{p}=2$ are
presented in Fig.\ref{basins1}. There are two attractors of
the system (limit cycles $r=10$ and $r\prime =\sqrt{50}$) and
their basins of attraction are marked by different colours, the
yellow area marks the basin of attraction of the attractor
$(\mathit{Re}\,a_{1})^{2}+(\mathit{Im}\,a_{1})^{2}=100$, whereas
the blue one refers to the basin of attraction of the attractor
$(\mathit{Re}\,a_{1})^{2}+(\mathit{Im}\,a_{1})^{2}=50$. Both 
basins have interesting geometries. The basin
corresponding to the circle of the radius $r\prime =\sqrt{50}$ has a
spiral-like form with the slip and width decreasing when
moving away from the centre. The remaining area (blue colour)
refers to the basin of attraction of the second attractor (the
circle of the radius $r=10$). Both attractors have a special
property. They are localised in such a way, that each
attractor is located partly in its own basin of attraction and
partly in the basin of attraction of the other attractor. As a
result - if the phase point starts from the part of the attractor
situated in the basin of attraction of the other one, it escapes to the
other attractor. However, if it starts from the part of the attractor
situated in its own basin of attraction, it does not change the
attractor. In analogy to semistable orbits these attractors
can be called the {\em semistable attractors}. So, the system with
Kerr nonlinearity is tuneable: an adequate choice of the initial
condition can result in the transition of the phase point from the
one attractor to the other. This property seems to be useful in
applications in optical switches.

\subsection{Parameters detuning}

\begin{figure}
\begin{center}
\includegraphics[width=8cm,height=9cm,angle=270]{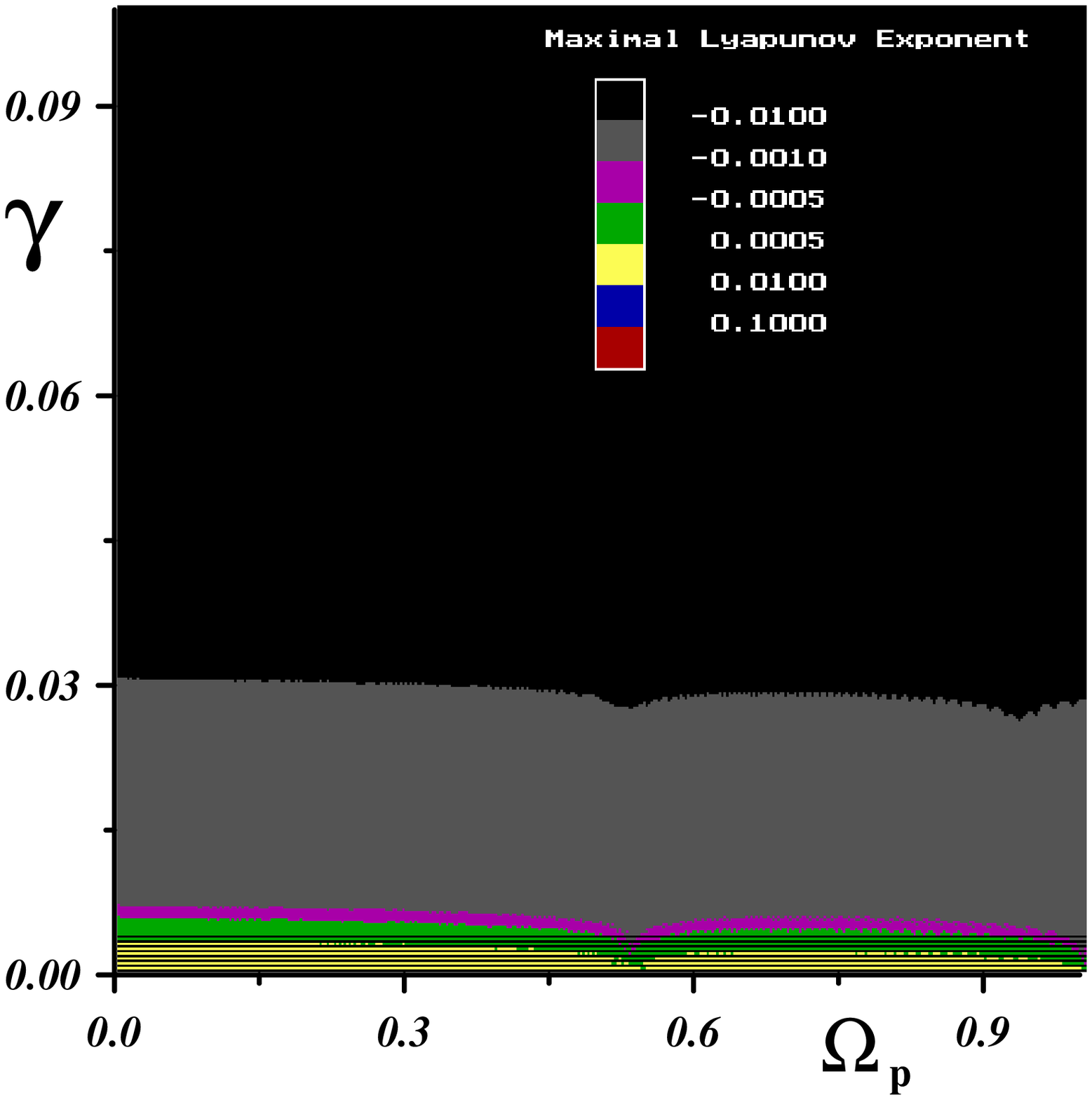}
\caption{Lyapunov map - the values of the maximal Lyapunow exponent
$\lambda_{1}$ for the parameters of the system (11): $\omega =1$,
$\epsilon =0.01$ and for the pump field amplitude $F=5$.
 The initial condition is: $\mathit{Re}\,a=10$, $\mathit{Im}\,a=0$. 
In the parameter space ($\Omega_{p}$, $\gamma$) the colours correspond 
to appropriate values of $\lambda_{1}$.} \label{lap1}
\end{center}
\end{figure}

\begin{figure}
\begin{center}
\includegraphics[width=9cm,height=9cm,angle=270]{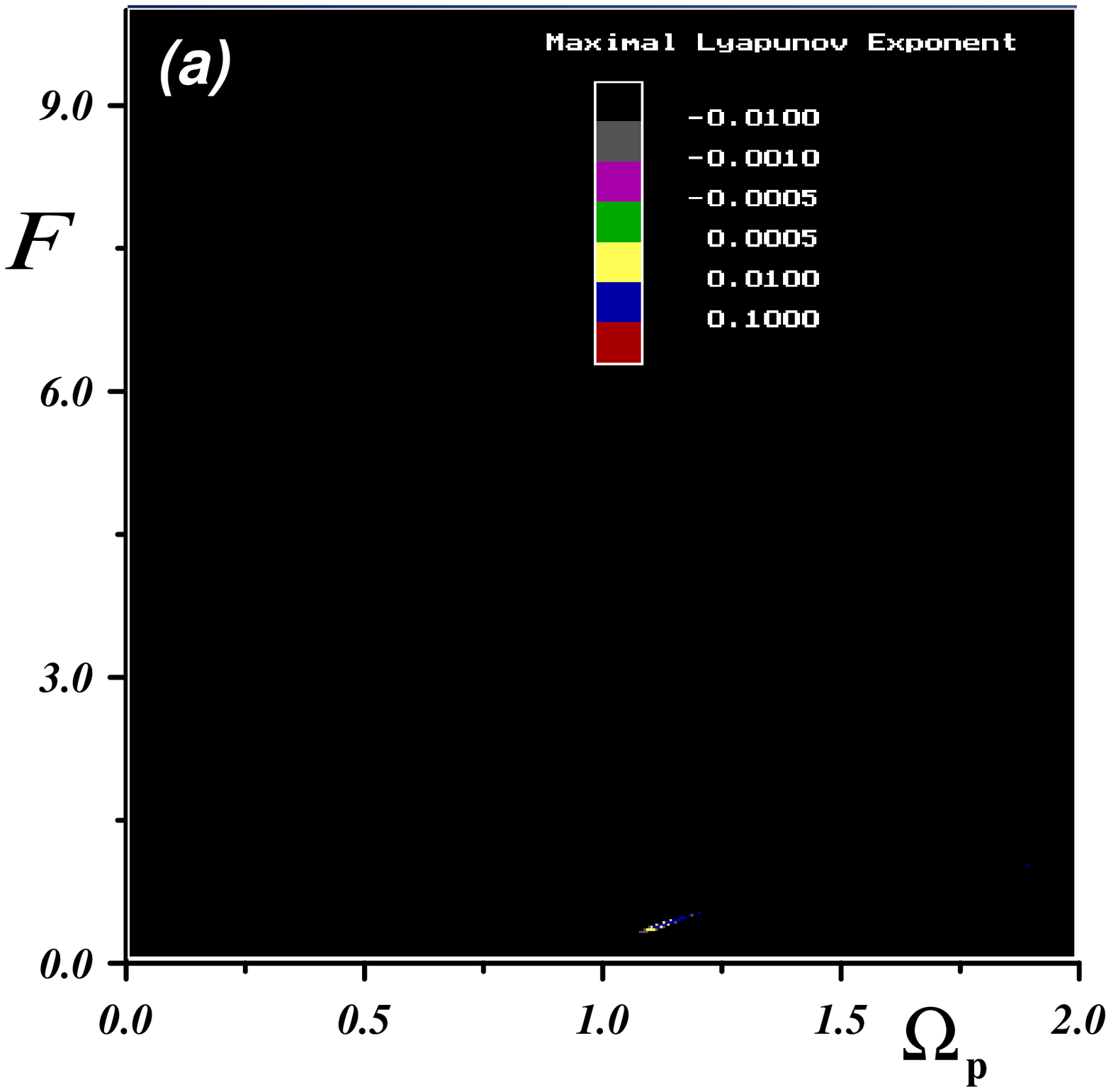}
\includegraphics[width=9cm,height=9cm,angle=270]{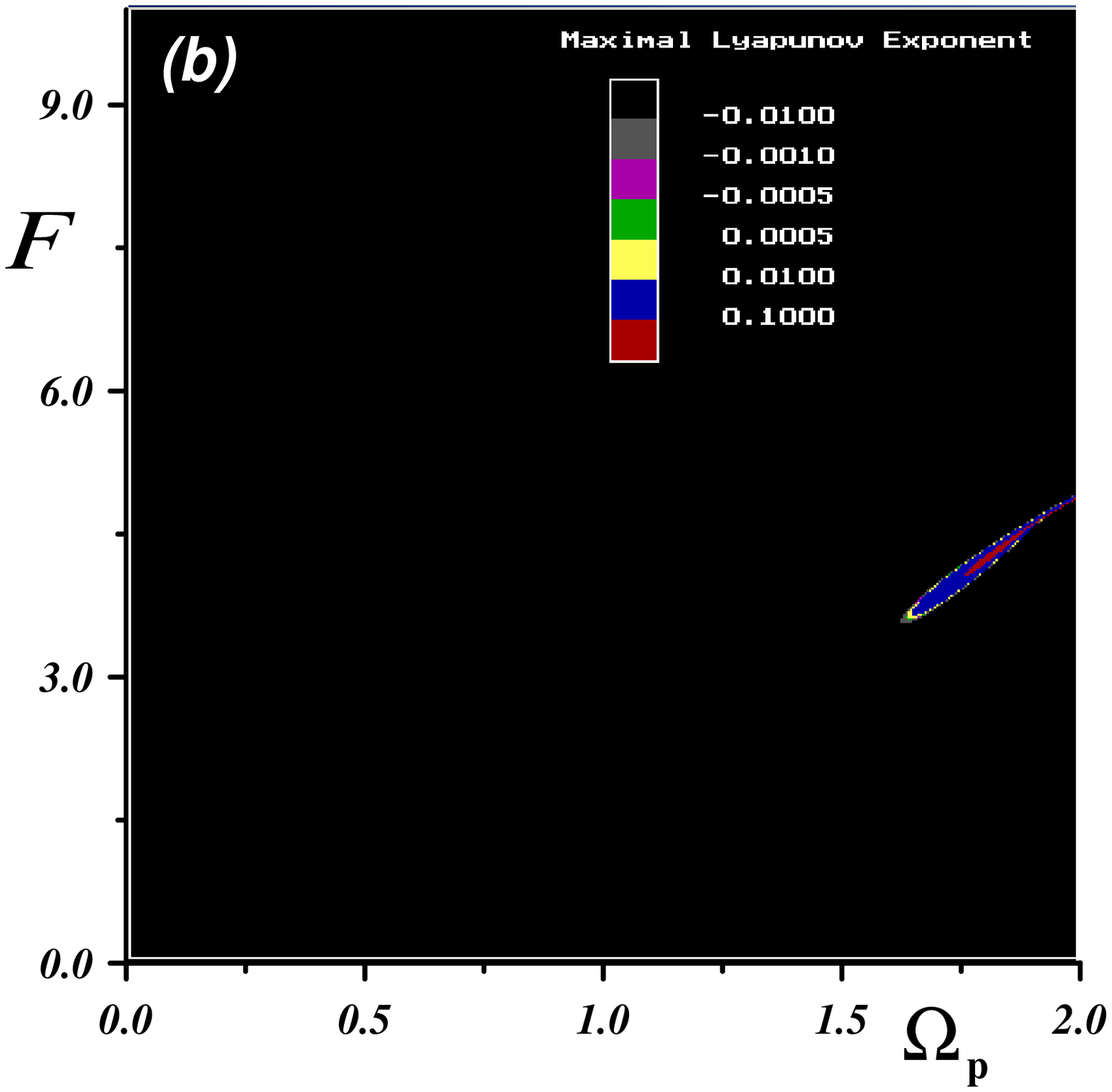}
\caption{Lyapunov maps - the values of the maximal Lyapunow exponent
$\lambda_{1}$ for the parameters of the system (11): $\omega =1$,
$\epsilon =0.01$ and for the initial condition: $\mathit{Re}\,a=10$, 
$\mathit{Im}\,a=0$. The parameter spaces ($\Omega_{p}$, $F$) 
are presented for (a) $\gamma=0.1$ and
(b) $\gamma=0.5$. For higher values of $\gamma$ only 
the periodic behaviour of the system is observed.} \label{lap23}
\end{center}
\end{figure}

One of the important properties of dynamical systems is their
sensitivity to the change in the system's parameters. A small
change in a parameter can lead to radical changes in the dynamics of
the system. This feature is frequently used to control the
dynamical systems. Globally, the behaviour of the system can
be shown on the so-called Lyapunov map in a parametric
space. We used here the well known procedure ~\cite{wolf} 
for numerical calculation of Lyapunov exponents (Lyapunov
spectrum $\lambda_{i}, i=1,2$). In Fig.\ref{lap1} we 
show the map of maximal Lyapunow
exponent $\lambda_{1}$ for the parameters of the system $\omega
=1$, $\epsilon =0.01$ and for the pump field amplitude $F=5$. The 
map is presented in the parameters space ($\Omega_{p}$, $\gamma$). 
The highest values of $\lambda_{1}$ corresponding to chaotic oscillations 
are marked by red and blue colours. The chaotic motion
exists only for weak damping. For higher damping 
constants we find only single islands of chaotic motion in the 
pump field parametric space ($\Omega_{p}$, $F$) (see Fig.\ref{lap23}).

\begin{figure}
\begin{center}
\includegraphics[width=6.5cm,height=6.5cm,angle=0]{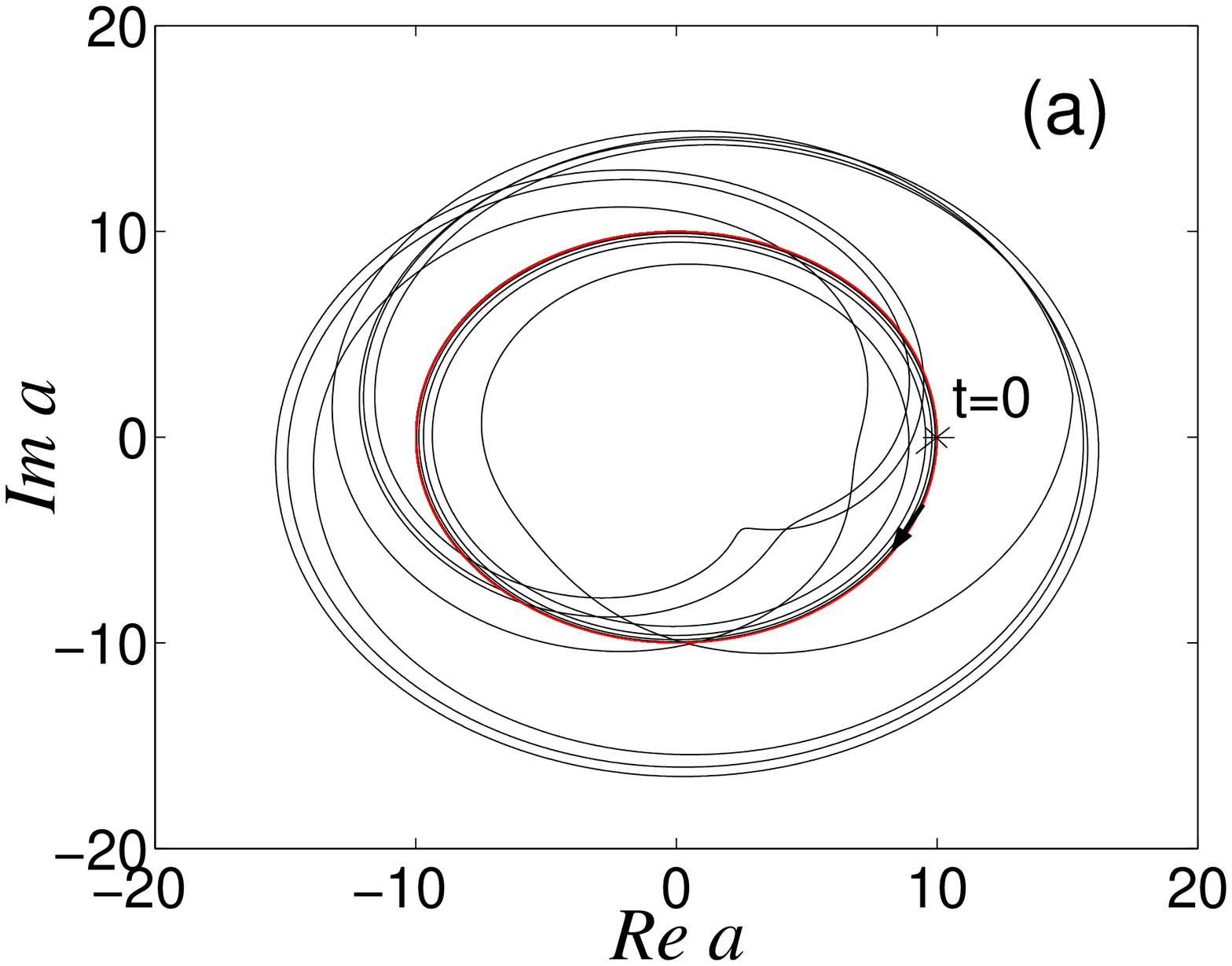}
\includegraphics[width=6.5cm,height=6.5cm,angle=0]{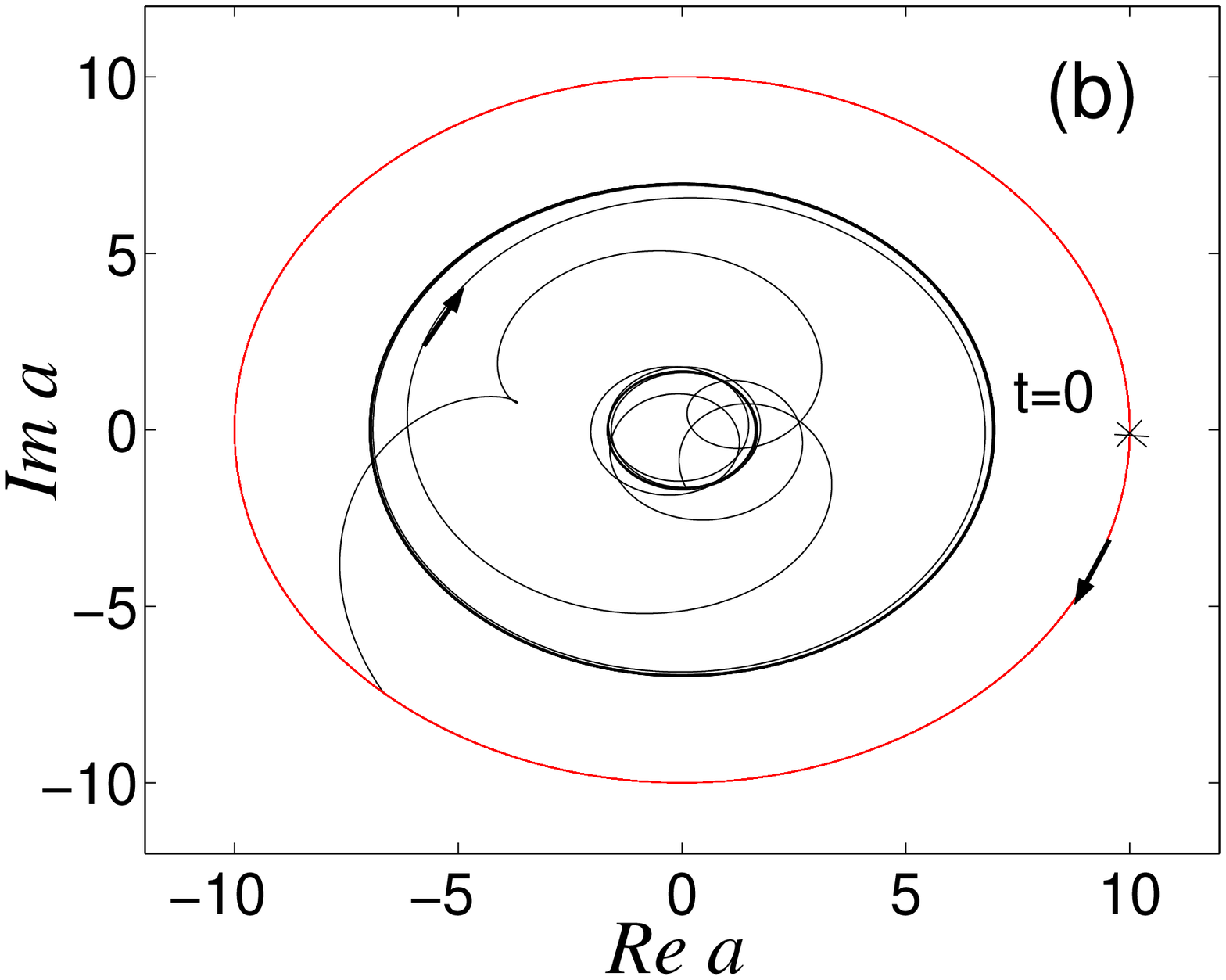}
\includegraphics[width=6.5cm,height=6.5cm,angle=0]{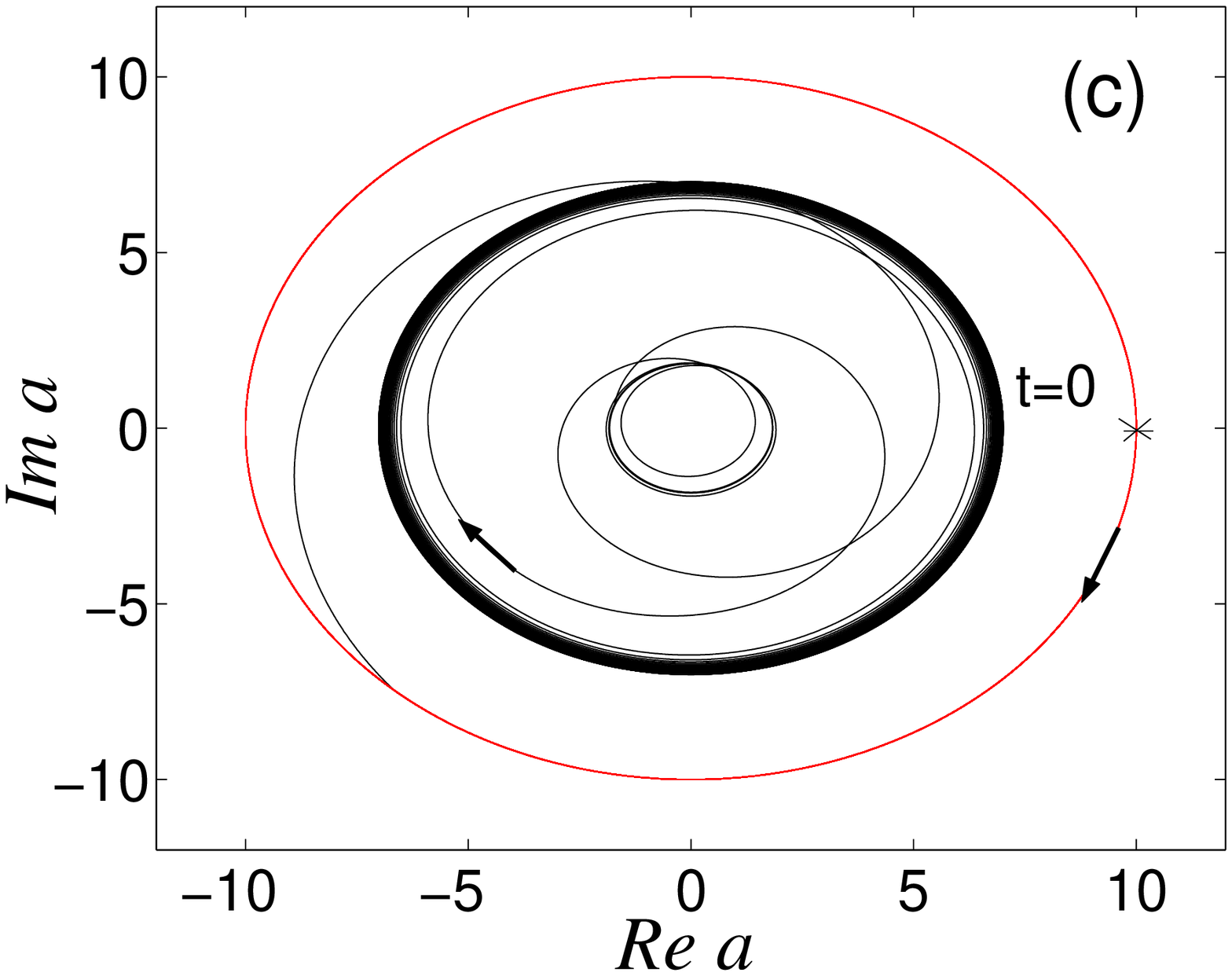}
\caption{The phase trajectories of the system (\ref{ker3}) for the
initial conditions: $\mathit{Re}\,a=10$, $\mathit{Im}\,a=0$ and for 
the following values of the system's parameters:
(a) $\omega =1$, $\epsilon =0.01$, $F=5$, $\Omega_{p}=2$ and
$\gamma =0.5$ for $0<t<20$, $\gamma =0.02$ for $20<t<40$ and
$\gamma =0.5$ for $40<t<60$; (b) $\omega =1$, $\gamma =0.5$,
$\epsilon =0.01$, $F=5$ and $\Omega_{p}=2$ for $0<t<20$,
$\Omega_{p} =4$ for $20<t<40$ and $\Omega_{p}=2$ for $40<t<100$;
(c) $\omega =1$, $\gamma =0.5$, $\epsilon =0.01$, $\Omega_{p}=2$
and $F=5$ for $0<t<20$, $F =2$ for $20<t<40$ and $F=5$ for
$40<t<250$.}
\label{odst}
\end{center}
\end{figure}

To control our system of Kerr oscillators we first change the value
of one of the parameters ($\gamma $, $F$ or $\Omega_{p}$) of the
system (\ref{ker2}) at time $t_{1}$. In such a way, the phase
point being in one of the two orbits (attractors) shown in
Fig.\ref{basins1} escapes from the attractor to the transient
state and sometimes to the new periodic state. Then after returning at 
 time $t_{2}>t_{1}$ to the initial values of this parameter, 
the additional periodic state disappears and the phase point trajectory 
tends to one of the two attractors of the system, depending 
on which basin of attraction was the phase point at time $t_{2}$ .  
So, through the appropriate choices of times $t_{1}$ end $t_{2}$ as well 
as the values of the parameters of the system we can control its
evolution. In particular, we can switch the system between two
stable periodic states. Such situations are illustrated in Fig.
\ref{odst}(a)--(c).
In Fig.\ref{odst}(a) the phase point starting from the orbit of the
radius $r=10$ (marked in red) after detuning the value of the
damping constant $\gamma$ in time $t_{1}=20$ from $\gamma
=0.5$ to $\gamma =0.02$ escapes through a transient state to 
the new periodic state $r=11,91$. After coming back with $\gamma $ 
to the initial value  $\gamma =0.5$ it goes into the initial orbit 
(of the radius $r=10$). In Fig. \ref{odst}(b)
the phase point starts also from the point lying on the orbit of the
radius $r=10$ and after detuning the parameter $\Omega_{p}$ from
$\Omega_{p}=2$ to $\Omega_{p}=4$ at time $t_{1}=20$ it 
escapes through a transient state to the new periodic state $r=1,84$. 
Then after coming back with the parameter $\Omega_{p}$ at time 
$t_{2}=40$ to the initial value
$\Omega_{p}=2$ it goes into the orbit of the radius $r\prime
=\sqrt{50}$ - the case of switching between the periodic orbits. 
Moreover, Fig.\ref{odst}(c) shows the phase
point starting from the site lying on the orbit of the radius $r=10$
after detuning the value of the parameter $F$ from $F=5$ to $F=2$
at time $t_{1}=20$ also moves to another periodic orbit $r=1,84$, 
and then after returning at time $t_{2}=40$ to the initial value 
of parameter $F$ ($F=5$) it goes into the orbit of the radius 
$r\prime =\sqrt{50}$. Generally, it is very interesting that if 
we change the parameters of the system, some extra periodic orbits 
appear.

\subsection{Generation of chaotic beats}

Since the publication of ~\cite{KG_PS}, the new type of signals
called "chaotic beats" has been investigated ~\cite{sliwa}
and experimentally generated ~\cite{cafagna,cafagna4}. There are two basic
kinds of chaotic beats: (1) the signals with chaotic envelopes and a
stable fundamental frequency, and (2) the signals with almost regular
collapses and revivals with small chaotic perturbations.
Generally, when the system is subjected to an external field, we
can generate the chaotic beats in two ways by modulation of the
amplitude or frequency of the external field.

The more effective method of generation of chaotic beats in
system (\ref{ker2}) seems to be the frequency
modulation, according to the formula: 

$\Omega_{p}\rightarrow\Omega_{p}(1+\Delta \Omega_{p}\sin (\mu t))$, 

where $\mu $ is the
frequency of modulation parameter, and $\Delta \Omega_{p} $ is the
amplitude of this modulation. Then the equation of motion for
variable $a$ takes the form:
\begin{equation}
\frac{da}{dt} = -i\omega a-i\epsilon a^{*}a^{2}-\gamma a
+Fe^{-i\Omega_{p}(1+\Delta \Omega_{p}\sin (\mu t))t}\;.
\label{ker3g}
\end{equation}
In numerical calculations we put $\Delta \Omega_{p}=0.1$.

\begin{figure}
\begin{center}
\includegraphics[width=8cm,height=12cm,angle=270]{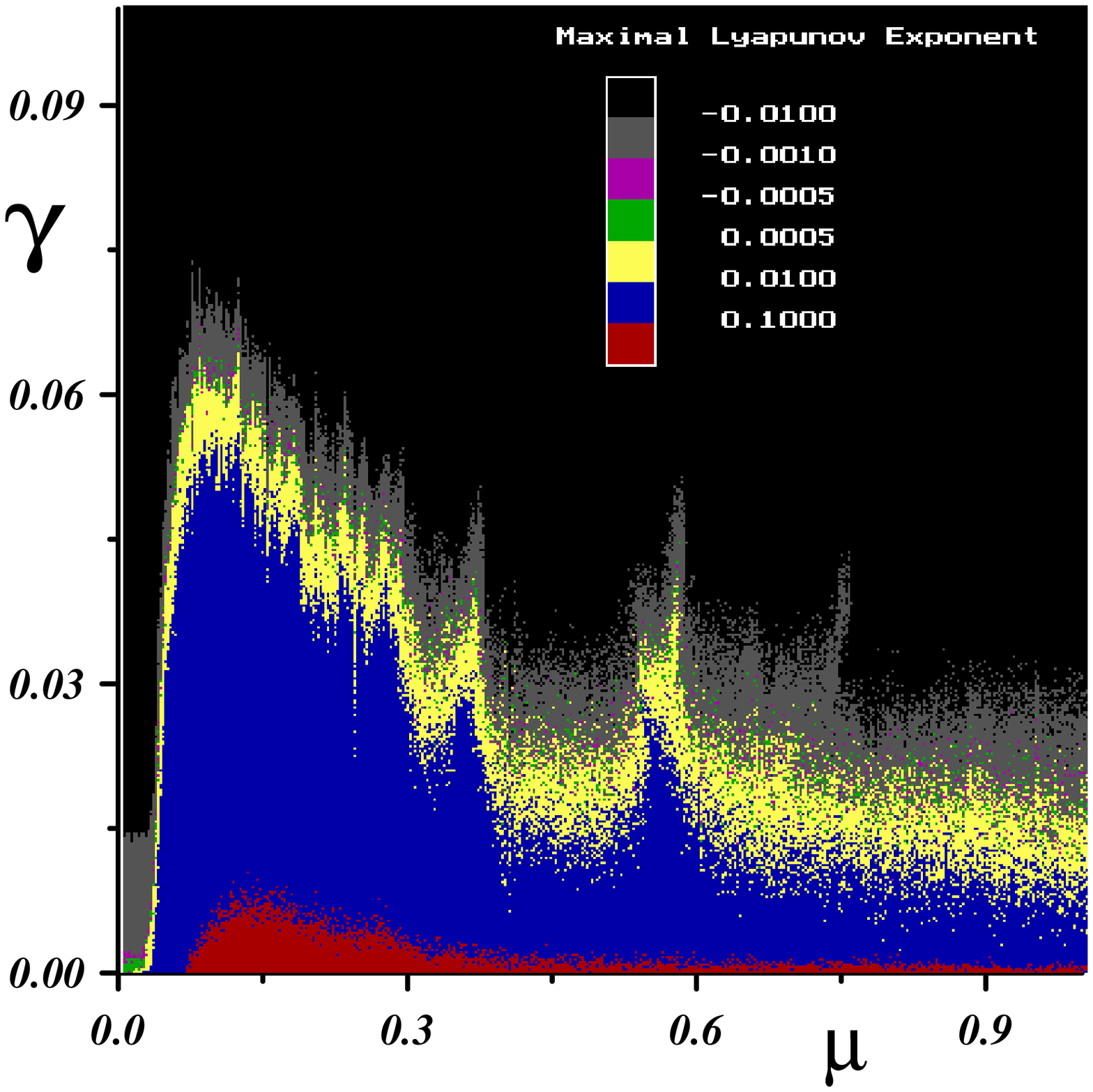}
\caption{The values of maximal Lyapunov exponent $\lambda_{1}$ for
the system (\ref{ker3g}) as a function of the damping parameter
$\gamma $ and the frequency modulation $\mu $, for $\epsilon
=0.01$, $F=5$, $\Omega_{p}=2$ and with the initial condition:
$\mathit{Re}\,a=10$, $\mathit{Im}\,a=0$. } \label{map_nres}
\end{center}
\end{figure}

\begin{figure}
\begin{center}
\includegraphics[width=8cm,height=4cm,angle=0]{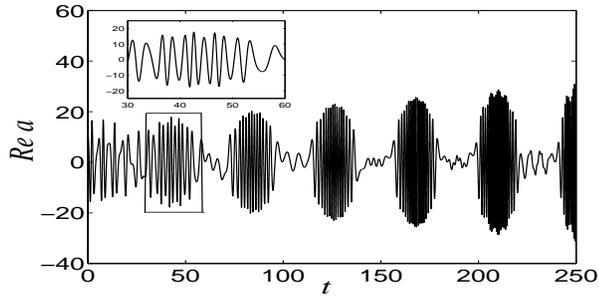}
\caption{Time dependence of $\mathit{Re}\,a$ of the system
(\ref{ker3g}), for $\gamma =0.02$  and $\mu =0.15$. The other
parameters are: $\omega =1$, $F=5$, $\epsilon =0.01$
and $\Omega_{p}=2$. The system starts from the initial condition:
$\mathit{Re}\,a=10$, $\mathit{Im}\,a=0$. Chaotic beats.}
\label{modnr02}
\end{center}
\end{figure}

The global dynamics of system (\ref{ker3g}) is presented in
Fig. \ref{map_nres} as a Lyapunov map in the parametric space of 
$(\gamma ,\mu)$, where the values of the first Lyapunov exponent
$\lambda_{1}$ are marked by appropriate colours. The highest
values of $\lambda_{1}$ corresponding to chaotic oscillations are
marked by red and blue colours. As we can see, they are
concentrated in the lower part of the map which corresponds to the low
values of the damping parameter $\gamma $ (mainly for $\gamma <0.05$). 
For higher $\gamma$ the map is dominated by the black 
and grey colours corresponding to the periodic states.
An example of chaotic beats generated in the system (\ref{ker3g}) 
through frequency modulation (with $\gamma =0.02$ and $\mu =0.15$) is 
presented in Fig.\ref{modnr02}. The spectrum of the  Lyapunov exponents
$\{0.0519,-0.0813\}$ of that system with the positive value of 
$\lambda_{1}$ indicates chaotic behaviour.

\begin{figure}
\begin{center}
\includegraphics[width=8cm,height=12cm,angle=270]{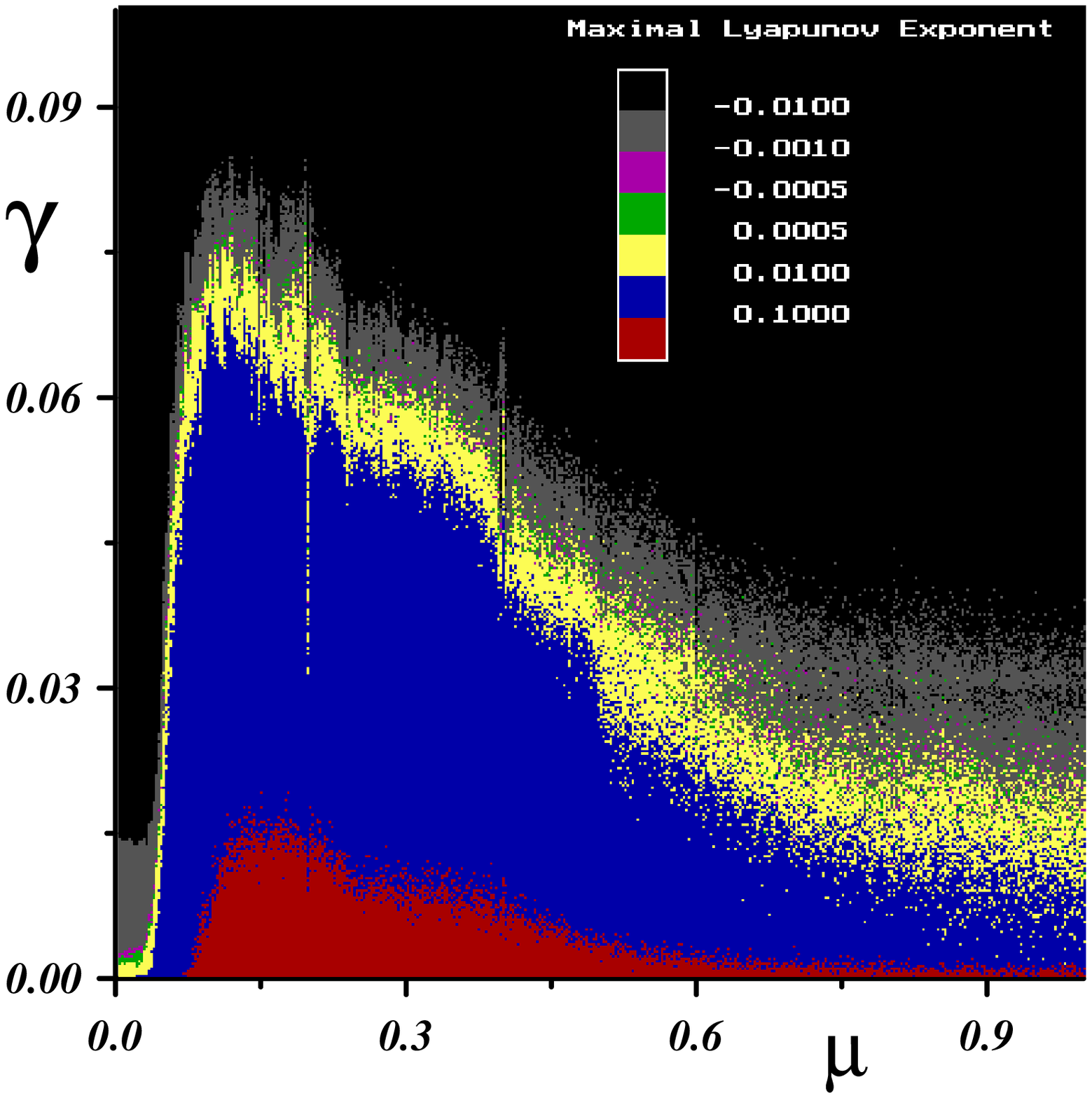}
\caption{The values of the maximal Lyapunov exponent $\lambda_{1}$
of system (\ref{ker3g}) as a function of the damping parameter
$\gamma $ and the frequency of modulation $\mu $, for $\epsilon
=0.01$, $F=5$ and $\Omega_{p}=\omega =1$ (the resonant case) and with 
the initial conditions: $\mathit{Re}\,a=10$, $\mathit{Im}\,a=0$.
}\label{map_res}
\end{center}
\end{figure}

Similar results were obtained for the resonance case
($\Omega_{p}=\omega $). The values of maximal Lyapunov exponent
$\lambda_{1}$ for system (\ref{ker3g}) as a function of the 
damping parameter $\gamma $ ($0<\gamma <0.5$) and the
frequency modulation $\mu $ ($0<\mu <1$) and for
$\Omega_{p}=\omega =1$ are presented in Fig.\ref{map_res}.

\begin{figure}
\begin{center}
\includegraphics[width=8cm,height=4cm,angle=0]{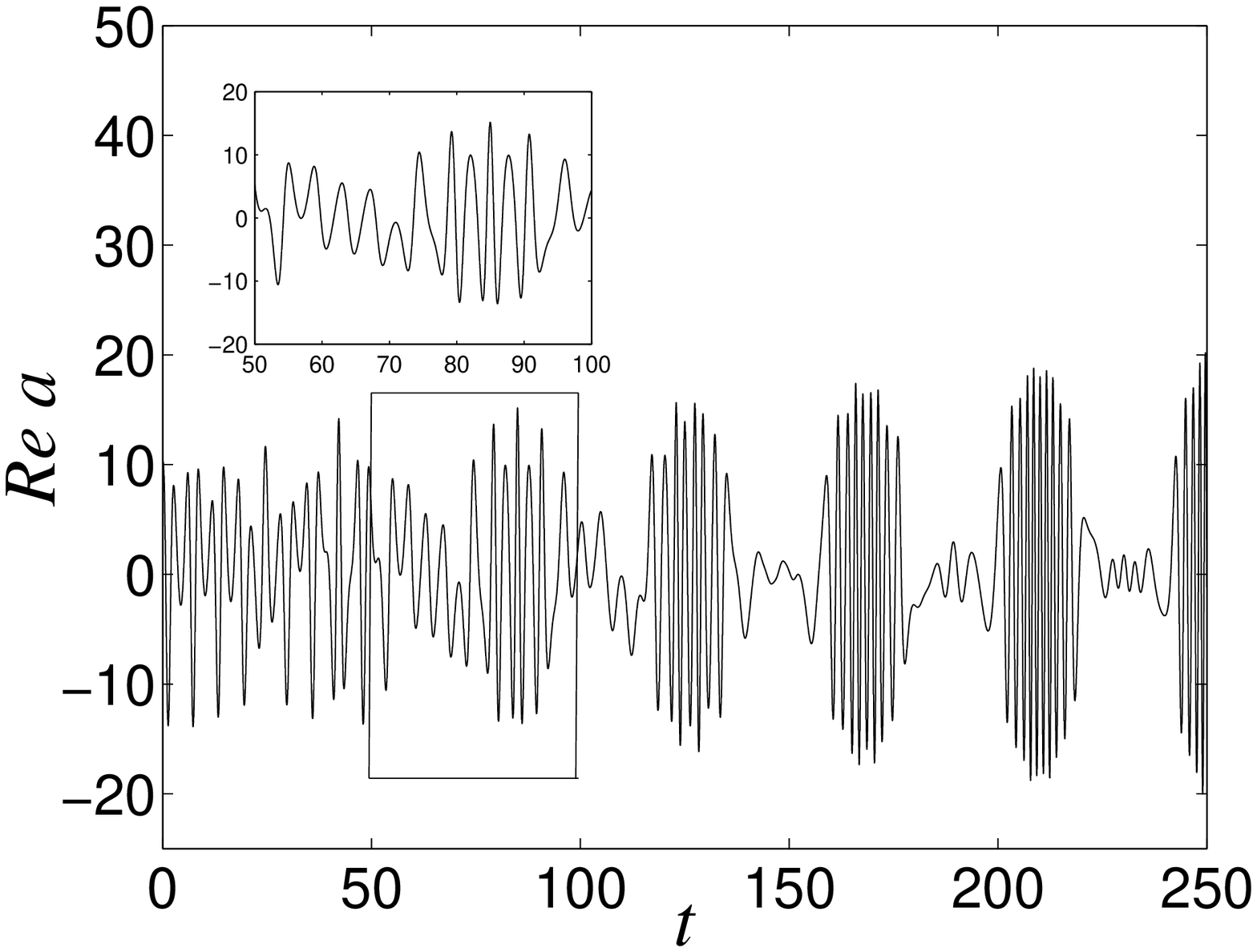}
\caption{Time dependence of $\mathit{Re}\,a$ of system
(\ref{ker3g}), for $\gamma =0.01$, $\mu =0.15$, $F=5$, $\epsilon
=0.01$ and $\Omega_{p}=\omega =1$. The system starts from the
initial condition: $\mathit{Re}\,a=10$,
$\mathit{Im}\,a=0$. Chaotic beats.}
\label{modr01}
\end{center}
\end{figure}

Analogically as for the non-resonance case, the chaotic behaviour
is obtained only for the low values of the damping parameter
$\gamma $ (mainly for $\gamma <0.05$). In Fig.\ref{modr01} we can see the
time dependence of $\mathit{Re}\,a$ of the system (\ref{ker3g}),
for $\gamma =0.01$, $\mu =0.15$ and $\Omega_{p}=\omega =1$.

The spectrum of Lyapunov exponents of the beats shown in
Fig.\ref{modr01} is $\{0.1085,-0.1230\}$ and contains positive value of
$\lambda_{1}$ indicating chaotic behaviour.

\section{The two coupled Kerr oscillators}

\subsection{Equations of motion}

It is interesting to know what happens to the dynamics of the
single Kerr oscillator after coupling it nonlinearly to another
analogous oscillator but of different frequency. Such a system of two
nonlinearly coupled Kerr oscillators (nonlinear couplers) is
described by the following hamiltonian:
\begin{equation}\label{ker4}
H=H_{0}+H_{1}+H_{2}\,,
\end{equation}
where:
\begin{eqnarray}
\label{H0} H_{0} & = & \sum_{j=1}^{2} \omega_{j}
 a_{j}^{*}a_{j}+\frac{1}{2}\sum_{j=1}^{2}\epsilon_{j}a_{j}^{*2}a_{j}^{2}\,,
 \\
 \label{H1}
 H_{1} & = & \epsilon_{12}a_{1}^{*}a_{2}^{*}a_{1}a_{2}\,, \\
 \label{H2}
 H_{2} & = & i\sum_{j=1}^{2}\left[F_{j}\left(a_{j}^{*}e^{-i\Omega_{jp}t}
  -a_{j}e^{i\Omega_{jp}t}\right)\right]\,.
\end{eqnarray}
The hamiltonian $H_{0}$ represents the two single Kerr
oscillators, $H_{1}$ is the hamiltonian of the interaction between
them and $H_{2}$ describes the interaction of the two oscillators
with the external fields. The values $a_{1}$ and $a_{2}$ are
complex dynamical variables, $\omega_{j}$ denote the frequencies
of the free vibrations of the two single oscillators;
$\epsilon_{1}$ and $\epsilon_{2}$ are Kerr parameters describing
nonlinearity in these sub-systems (oscillators) and
$\epsilon_{12}$ is the parameter of nonlinear coupling between
them.

The equations of motion for variables $a_{1}$ and $a_{2}$ have the
form:
\begin{eqnarray}\label{ker11a}
\frac{da_{1}}{dt} & = & -i\omega_{1}
a_{1}-i\epsilon_{1}a_{1}^{*}a_{1}^{2}-i\epsilon_{12}a_{2}^{*}a_{1}a_{2}+\\
&+&F_{1}e^{-i\Omega_{1p}t}-\gamma_{1}a_{1}\,, \nonumber \\
\label{ker11b} 
\frac{da_{2}}{dt} & = & -i\omega_{2}
a_{2}-i\epsilon_{2}a_{2}^{*}a_{2}^{2}-i\epsilon_{12}a_{1}^{*}a_{1}a_{2}+\\
&+&F_{2}e^{-i\Omega_{2p}t}-\gamma_{2}a_{2}\,,\nonumber
\end{eqnarray}
where the terms $\gamma_{1} a_{1}$ and $\gamma_{2} a_{2}$,
describing the dissipation of energy, with damping constants
$\gamma_{1}$ and $\gamma_{2}$, are added to the equations of
motion on phenomenological grounds. 
In the autonomous and conservative case, that is when
$\gamma_{1}=\gamma_{2}=F_{1}=F_{2}=0$ the solutions of equations
(\ref{ker11a})--(\ref{ker11b}) have the form:
\begin{eqnarray}\label{aut_s1}
a_{1}(t) & = & a_{10}e^{-i(\omega_{1}
+\epsilon_{1}a_{10}^{*}a_{10}+\epsilon_{12}a_{20}^{*}a_{20})t}\,, \\
\label{aut_s2} a_{2}(t) & = & a_{20}e^{-i(\omega_{2}
+\epsilon_{2}a_{20}^{*}a_{20}+\epsilon_{12}a_{10}^{*}a_{10})t}\,,
\end{eqnarray}
where $a_{1}(t)\mid_{t=0}\equiv a_{10}$ and
$a_{2}(t)\mid_{t=0}\equiv
a_{20}$ denote initial conditions.\\
In the nonautonomous case (\ref{ker11a})--(\ref{ker11b}) we find
periodic solutions in the form of the following functions (in analogy 
to the case of the single Kerr oscillator):
\begin{eqnarray}\label{naut_s1}
a_{1}(t) & = &
x_{1}e^{-i(\omega_{1}+\epsilon_{1}x_{1}^{*}x_{1}+\epsilon_{12}x_{2}^{*}x_{2})t}\,,\\
\label{naut_s2} a_{2}(t) & = &
x_{2}e^{-i(\omega_{2}+\epsilon_{2}x_{2}^{*}x_{2}+\epsilon_{12}x_{1}^{*}x_{1})t}\,,
\end{eqnarray}
that is in the form of the solutions of the autonomous one. 
Functions (\ref{naut_s1})--(\ref{naut_s2}) satisfy the equations
of motion (\ref{ker11a})--(\ref{ker11b}) on condition that
$x_{j}=F_{j}/\gamma_{j}$, $j=1,2$. As a result these solutions
have the form:
\begin{eqnarray}
\label{ker11ar} a_{1}(t) & = &
\frac{F_{1}}{\gamma_{1}}\exp\left[-i\left(\omega_{1} +\epsilon_{1}
\frac{F_{1}^{2}}{\gamma_{1}^{2}}+\epsilon_{12}\frac{F_{2}^{2}}
{\gamma_{2}^{2}}\right)t\right] \,,\\
\label{ker11br}a_{2}(t) & = &
\frac{F_{2}}{\gamma_{2}}\exp\left[-i\left(\omega_{2} +\epsilon_{2}
\frac{F_{2}^{2}}{\gamma_{2}^{2}}+\epsilon_{12}\frac{F_{1}^{2}}
{\gamma_{1}^{2}}\right)t\right]\,.
\end{eqnarray}
Then $\Omega_{1p}=\omega_{1}+\epsilon_{1}F_{1}^{2}/\gamma_{1}^{2}
+\epsilon_{12}F_{2}^{2}/\gamma_{2}^{2}$ and
$\Omega_{2p}=\omega_{2}+\epsilon_{2}F_{2}^{2}/\gamma_{2}^{2}
+\epsilon_{12}F_{1}^{2}/\gamma_{1}^{2}$.\\
In the phase plane
$(\mathit{Re}\,a_{1},\,\mathit{Im}\,a_{1},\,\mathit{Re}\,a_{2},\,\mathit{Im}\,a_{2})$
the periodic solutions (\ref{ker11ar})--(\ref{ker11br}) satisfy
the phase equations (limit cycles - circles):
\begin{equation}\label{okrag1}
(\mathit{Re}\,a_{j})^{2}+(\mathit{Im}\,a_{j})^{2}=
\frac{F_{j}^{2}}{\gamma_{j}^{2}}\,,\,\,\,\,\,\mbox{where $j=1,2$},
\end{equation}
for any values of frequencies $\omega_{1}$, $\omega_{2}$.

\subsection{The dynamics of the system in the phase space}

Coupling the single Kerr oscillator described by equation
(\ref{ker2}) to the other analogous oscillator but with different own
frequency causes distinct changes in its dynamics, as illustrated by
 the diagrams in the phase space.

\begin{figure}
\begin{center}
\includegraphics[width=7cm,height=7cm,angle=0]{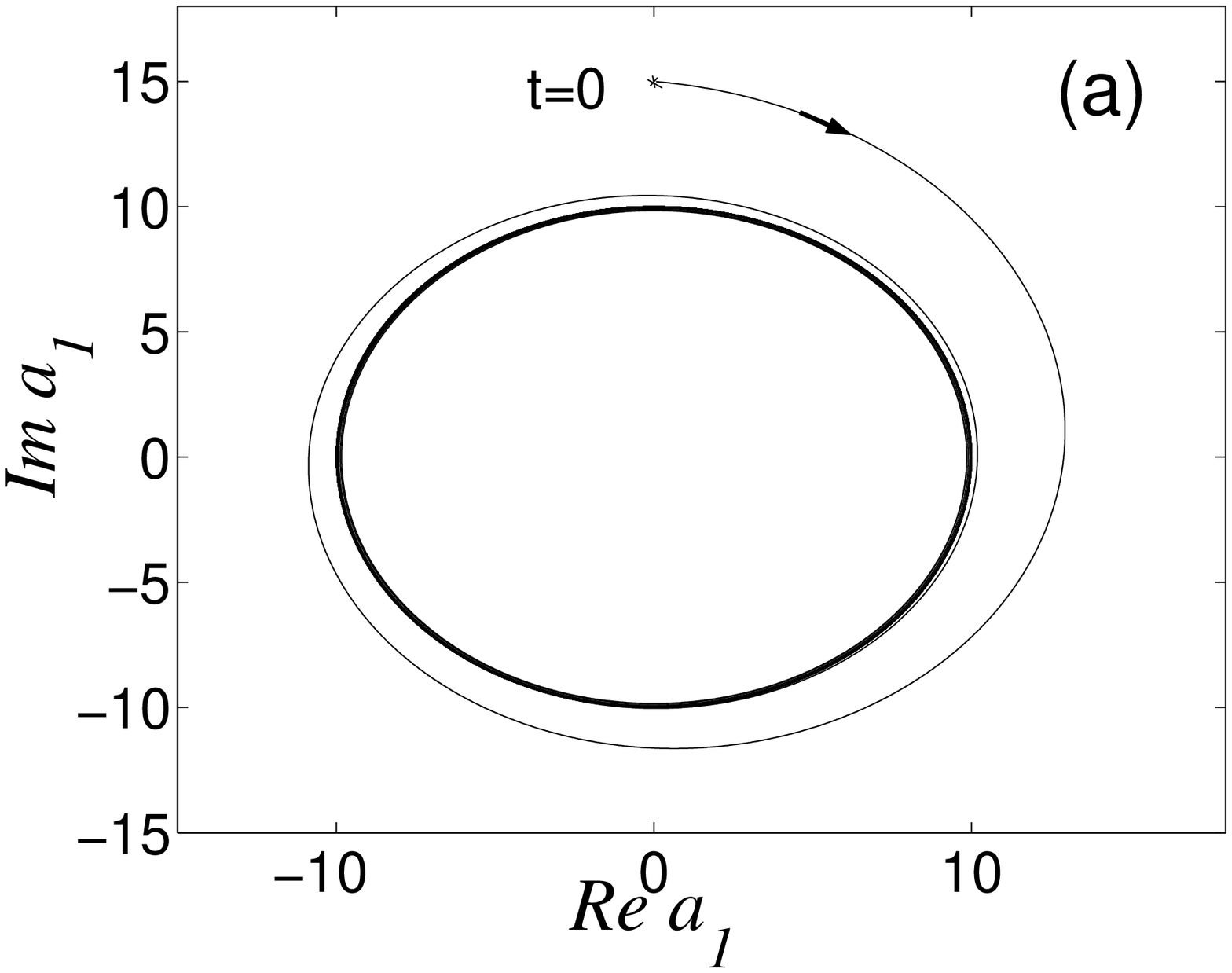}
\includegraphics[width=7cm,height=7cm,angle=0]{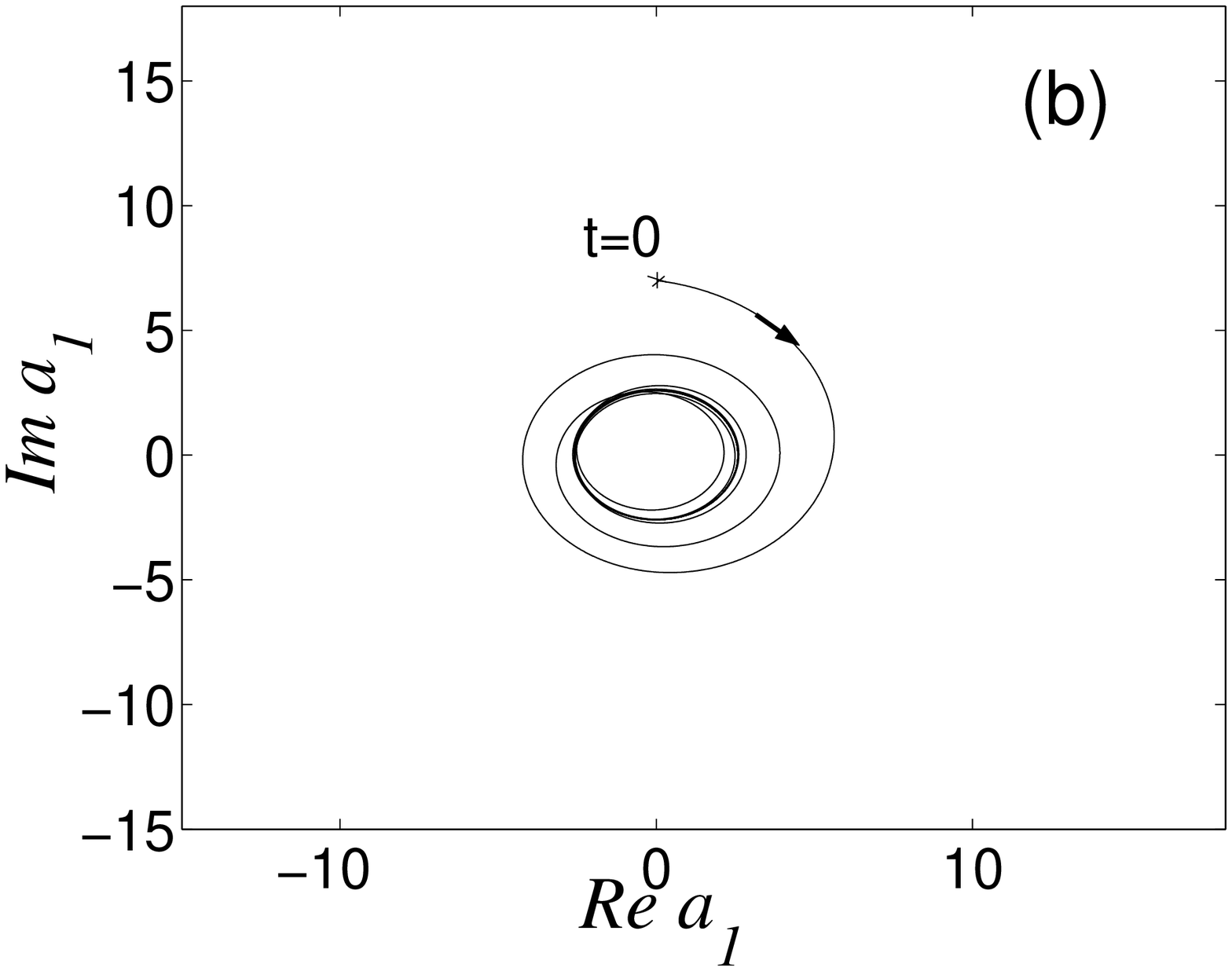}
\caption{The phase trajectories of sub-system (\ref{ker11a}) if
$\omega_{1}=1$, $\omega_{2}=0.5$,
$\epsilon_{1}=\epsilon_{2}=0.01$, $F_{1}=F_{2}=5$,
$\gamma_{1}=\gamma_{2}=0.5$, $\Omega_{1p}=3.0$, $\Omega_{2p}=2.5$,
$\epsilon_{12}=0.01$ and for the initial conditions: (a)
$\mathit{Re}\,a_{10}=0$, $\mathit{Im}\,a_{10}=15$,
$\mathit{Re}\,a_{20}=10$, $\mathit{Im}\,a_{20}=0$; (b)
$\mathit{Re}\,a_{10}=0$, $\mathit{Im}\,a_{10}=7$,
$\mathit{Re}\,a_{20}=10$, $\mathit{Im}\,a_{20}=0$.}\label{fig2}
\end{center}
\end{figure}

Let us consider two nonlinearly coupled Kerr oscillators
(sub-systems) described by equations (\ref{ker11a})--(\ref{ker11b})
if $\omega_{1}=1$, $\omega_{2}=0.5$,
$\epsilon_{1}=\epsilon_{2}=0.01$, $F_{1}=F_{2}=5$,
$\gamma_{1}=\gamma_{2}=0.5$, $\Omega_{1p}=3.0$, $\Omega_{2p}=2.5$
and $\epsilon_{12}=0.01$. Then, in compliance with
(\ref{ker11ar})--(\ref{ker11br}), the periodic solutions of
equations (\ref{ker11a})--(\ref{ker11b}) have the form:
\begin{eqnarray}\label{sprz_1}
a_{1}(t) & = & 10e^{-3.0it}\,, \\
\label{sprz_2} a_{2}(t) & = & 10e^{-2.5it}\,,
\end{eqnarray}
and in the phase space they satisfy the following equations:
\begin{eqnarray}\label{traj_1}
(\mathit{Re}\,a_{1})^{2}+(\mathit{Im}\,a_{1})^{2} & = & 100\,,
\\
\label{traj_2} (\mathit{Re}\,a_{2})^{2}+(\mathit{Im}\,a_{2})^{2} &
= & 100\,.
\end{eqnarray}
As a result, for the initial conditions $a_{10}=10$ and
$a_{20}=10$ the phase points of both subsystems draw the same
circle of the radius $r=10$ described by equations
(\ref{traj_1})--(\ref{traj_2}), with frequencies $\Omega_{1p}=3.0$
and $\Omega_{2p}=2.5$, respectively. But if the phase point
representing the first subsystem (at frequency $\omega_{1}=1$)
starts from another initial condition  instead of $a_{10}=10$, and
$a_{20}$ is fixed ($a_{20}=10$), the following behaviour is
observed: the phase point representing the first sub-system after
some time tends to one of the two orbits being attractors of
subsystem (\ref{ker11a}):
$(\mathit{Re}\,a_{1})^{2}+(\mathit{Im}\,a_{1})^{2}=100$ or
$(\mathit{Re}\,a_{1})^{2}+(\mathit{Im}\,a_{1})^{2}=6.6987$.
The second orbit belongs to the periodic solution:
$a_{1}(t)=(10/(1-i(2-\sqrt{3}))exp(-3.0it)$ and $a_{2}(t)=(10/(1-i(2-\sqrt{3}))exp(-2.5it)$. 
This situation is illustrated in Fig. \ref{fig2}.
In Fig. \ref{fig2}(a) the phase point representing the first
sub-system starts from the initial condition
$\mathit{Re}\,a_{10}=0$, $\mathit{Im}\,a_{10}=15$ (the initial
condition for the second subsystem (\ref{ker11b}) is fixed:
$\mathit{Re}\,a_{20}=10$, $\mathit{Im}\,a_{20}=0$) and after the
time $t=50$ it goes into the attractor of the radius $r=10$,
described by equation (\ref{traj_1}). However, Fig.\ref{fig2}(b) shows
the phase point starting from the initial
condition $\mathit{Re}\,a_{10}=0$, $\mathit{Im}\,a_{10}=7$ and going
into the orbit of the radius $r'=\sqrt{6.6987}$. 

The third solution of (16) and (17) is completely unstable and has the form:\\ 
$a_{1}(t)=(10/(1-i(2+\sqrt{3}))exp(-3.0it)$ \\ and $a_{2}(t)=(10/(1-i(2+\sqrt{3}))exp(-2.5it)$.
\subsection{Basins of attraction}

\begin{figure}
\begin{center}
\includegraphics[width=7cm,height=8cm,angle=270]{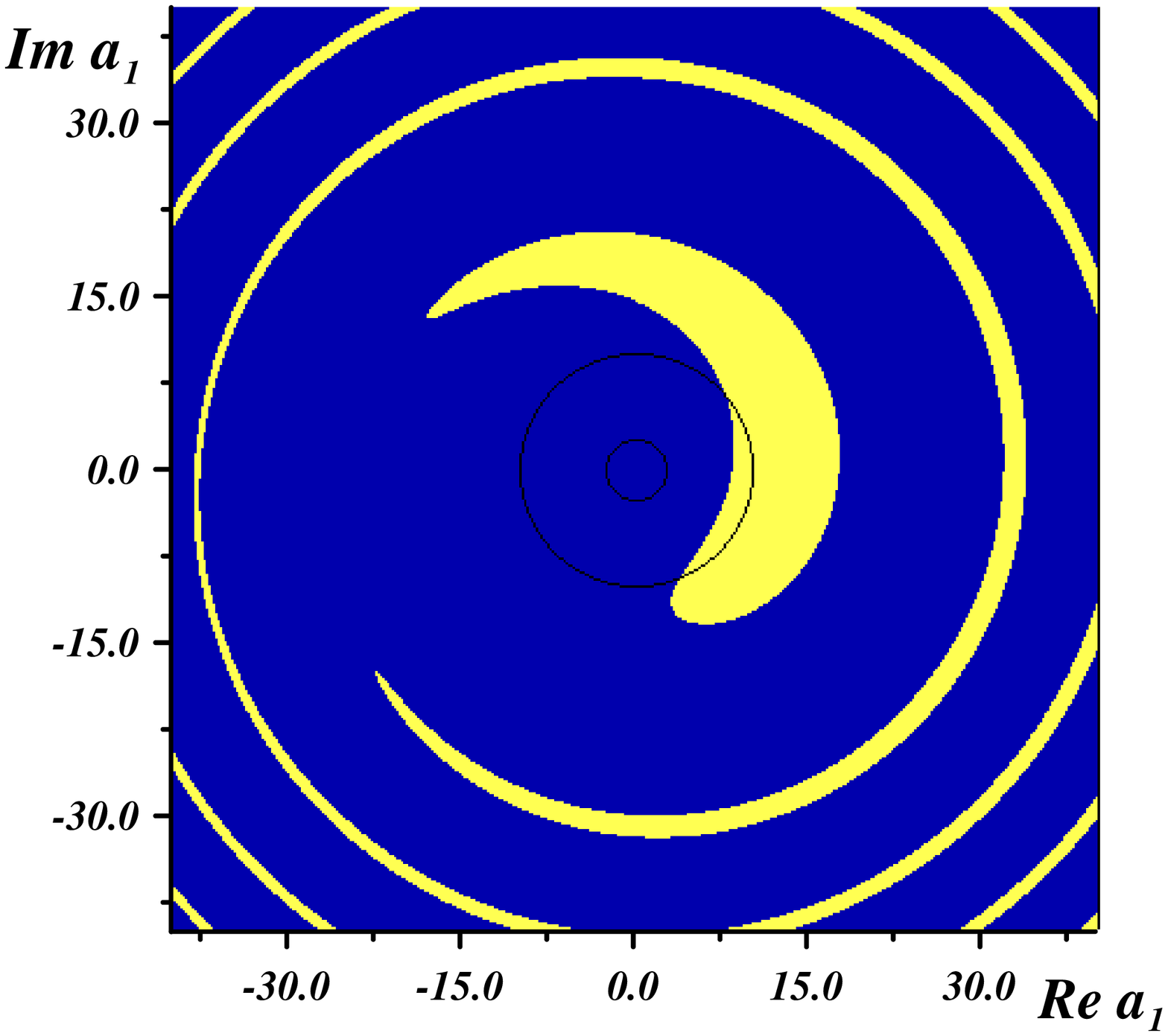}
\caption{Basins of attraction of two coupled Kerr oscillators.
The parameters of the systems are:
$\omega_{1}=1$, $\omega_{2}=0.5$,
$\epsilon_{1}=\epsilon_{2}=0.01$, $F_{1}=F_{2}=5$,
$\gamma_{1}=\gamma_{2}=0.5$, $\Omega_{1p}=3.0$, $\Omega_{2p}=2.5$,
$\epsilon_{12}=0.01$, and the initial conditions are:
$\mathit{Re}\,a_{10}=0$, $\mathit{Im}\,a_{10}=15$,
$\mathit{Re}\,a_{20}=10$, $\mathit{Im}\,a_{20}=0$. Stable
($r'=\sqrt{6.6987}$) and semistable ($r=10$) attractors.}
\label{basins2}
\end{center}
\end{figure}

Fig.\ref{basins2} shows two attractors and their basins of
attraction of sub-system (\ref{ker11a}) marked by 
appropriate colours; the initial condition of sub-system
(\ref{ker11b}) is fixed ($a_{20}=10$). The yellow area marks the
basin of attraction of the attractor of the radius $r=10$, whereas
the blue one corresponds to the basin of attraction of the attractor
of the radius $r' =\sqrt{6.6987}$. The basin marked in yellow has a special geometry: it
consists of two separate areas, one of which has a spiral-like
form. Both, the slip of the spiral and its width decrease towards
moving away from the centre, similarly, as for 
the single Kerr oscillator. Contrary to the case of the 
single Kerr oscillator there is an island in the central
part of the basin. The remaining area (blue colour) is the basin of
attraction of the other attractor (the circle of the radius 
$r'=\sqrt{6.6987}$).

The attractor of the radius $r' =\sqrt{6.6987}$ is stable (it is
fully in its own basin of attraction), however the
attractor of the radius $r=10$ is semistable (it is partly in
its own basin of attraction and partly in the basin of attraction
of the other attractor). As a result the transition from the
attractor of the radius $r=10$ to the attractor of the radius 
$r'=\sqrt{6.6987}$ is possible, but that in opposite direction is
impossible, because the phase point starting from any position on
the circle $(\mathit{Re}\,a)^{2}+(\mathit{Im}\,a)^{2}=6.6987$ always
returns to it.

\begin{figure}
\begin{center}
\includegraphics[width=7cm,height=8cm,angle=270]{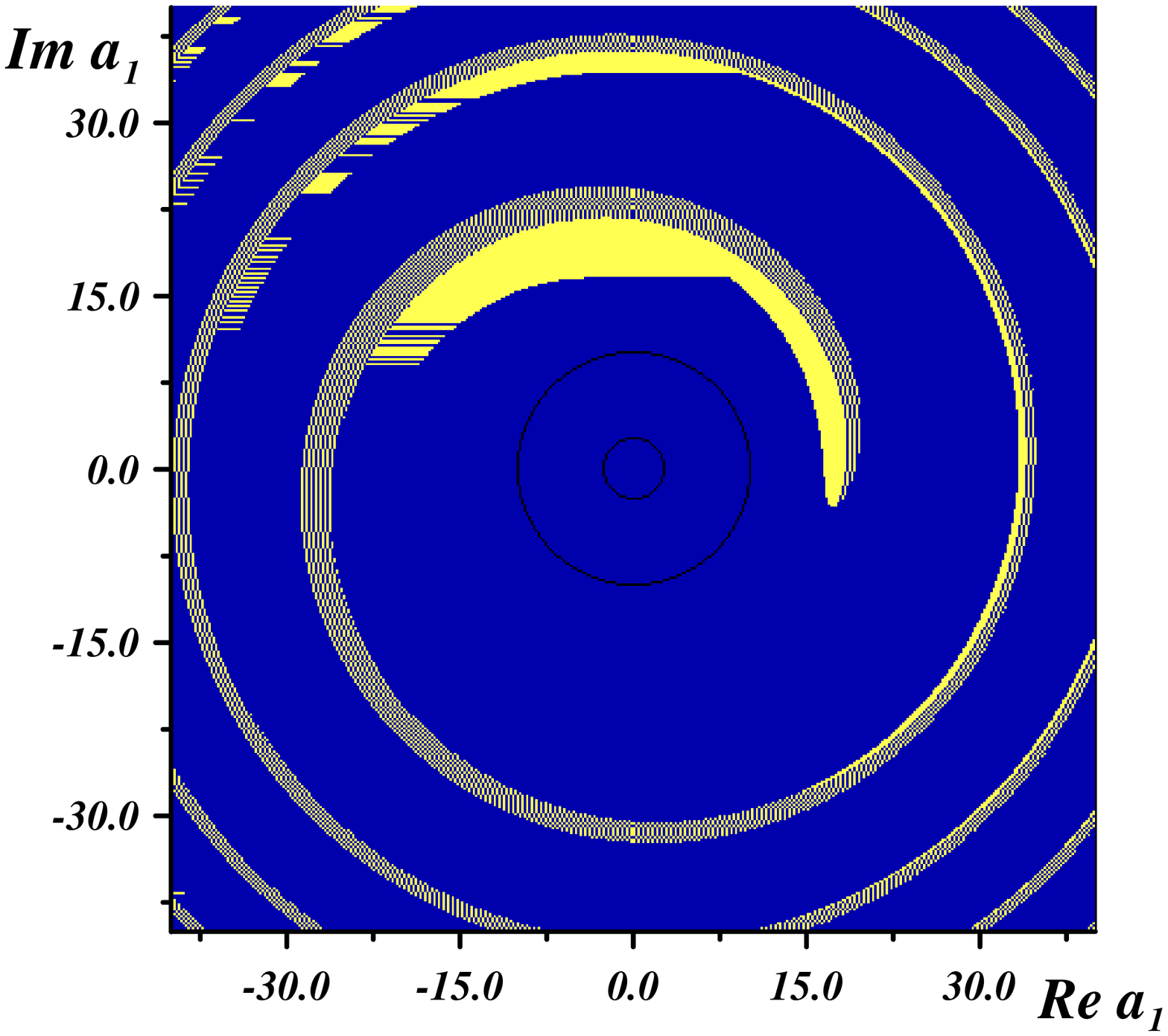}
\caption{Basins of attraction of two coupled Kerr oscillators.
The parameters are the same as in Fig. \ref{basins2}, 
but $\mathit{Re}\,a_{20}=0$, $\mathit{Im}\,a_{20}=0$. Stable 
($r'=\sqrt{6.6987}$) and unstable ($r=10$) attractors. } 
\label{basins2a}
\end{center}
\end{figure}

The types of attractors change after changing the
initial condition of the sub-system (\ref{ker11b}). For example, 
Fig.\ref{basins2a} shows the attractors and their basins of
attractions of sub-system (\ref{ker11a}) for $a_{20}=0$. The
basin of attraction corresponding to the circle of the radius $r=10$
is marked by yellow; the remaining area (blue colour) refers to the
basin of attraction of the circle of $r'=\sqrt{6.6987}$. 
As we can see, in this case the attractor with radii
($r' =\sqrt{6.6987}$) is stable, and the other one is
completely unstable.
\begin{figure}
\begin{center}
\includegraphics[width=8cm,height=8cm,angle=270]{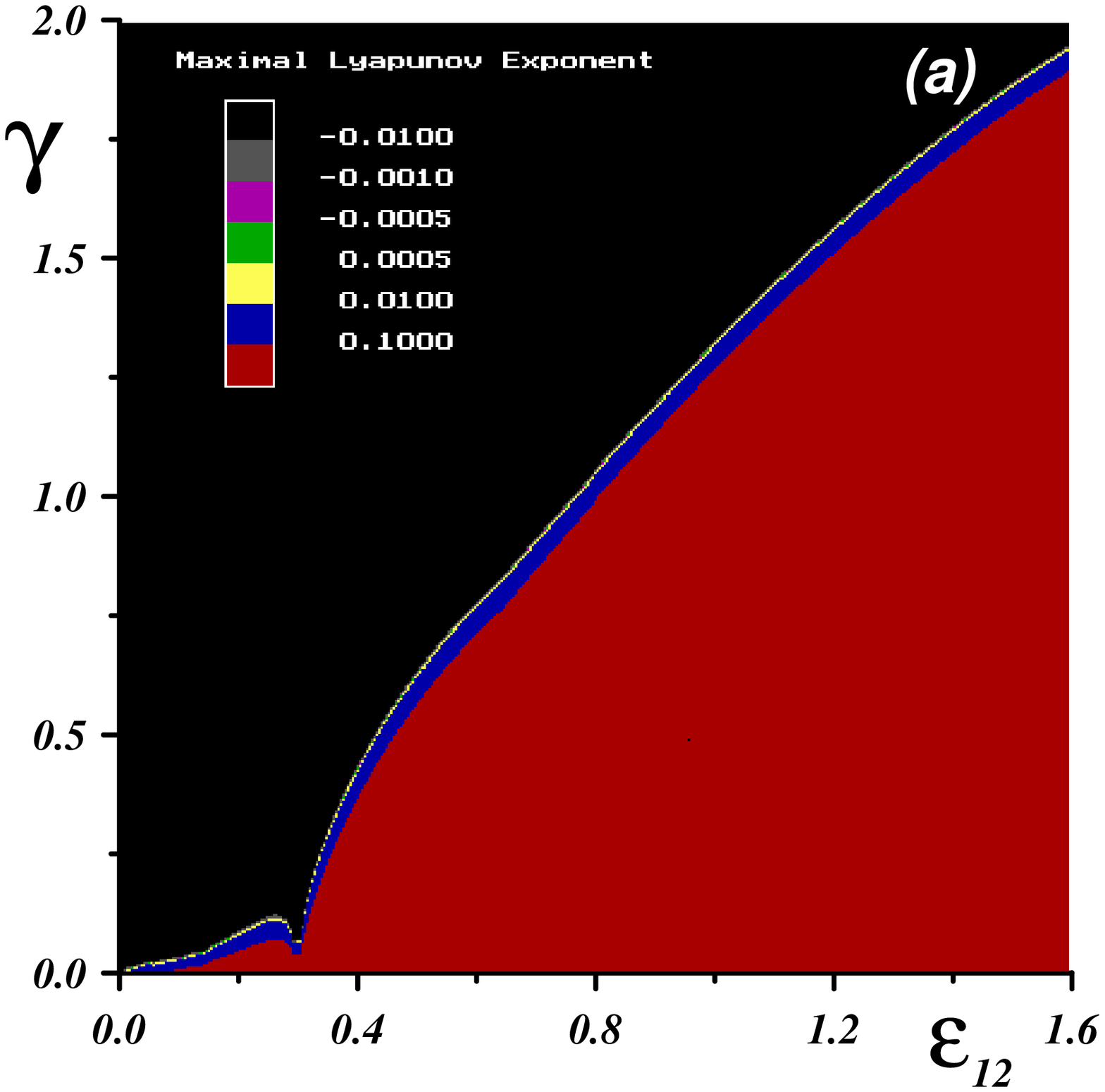}
\includegraphics[width=7cm,height=8cm,angle=270]{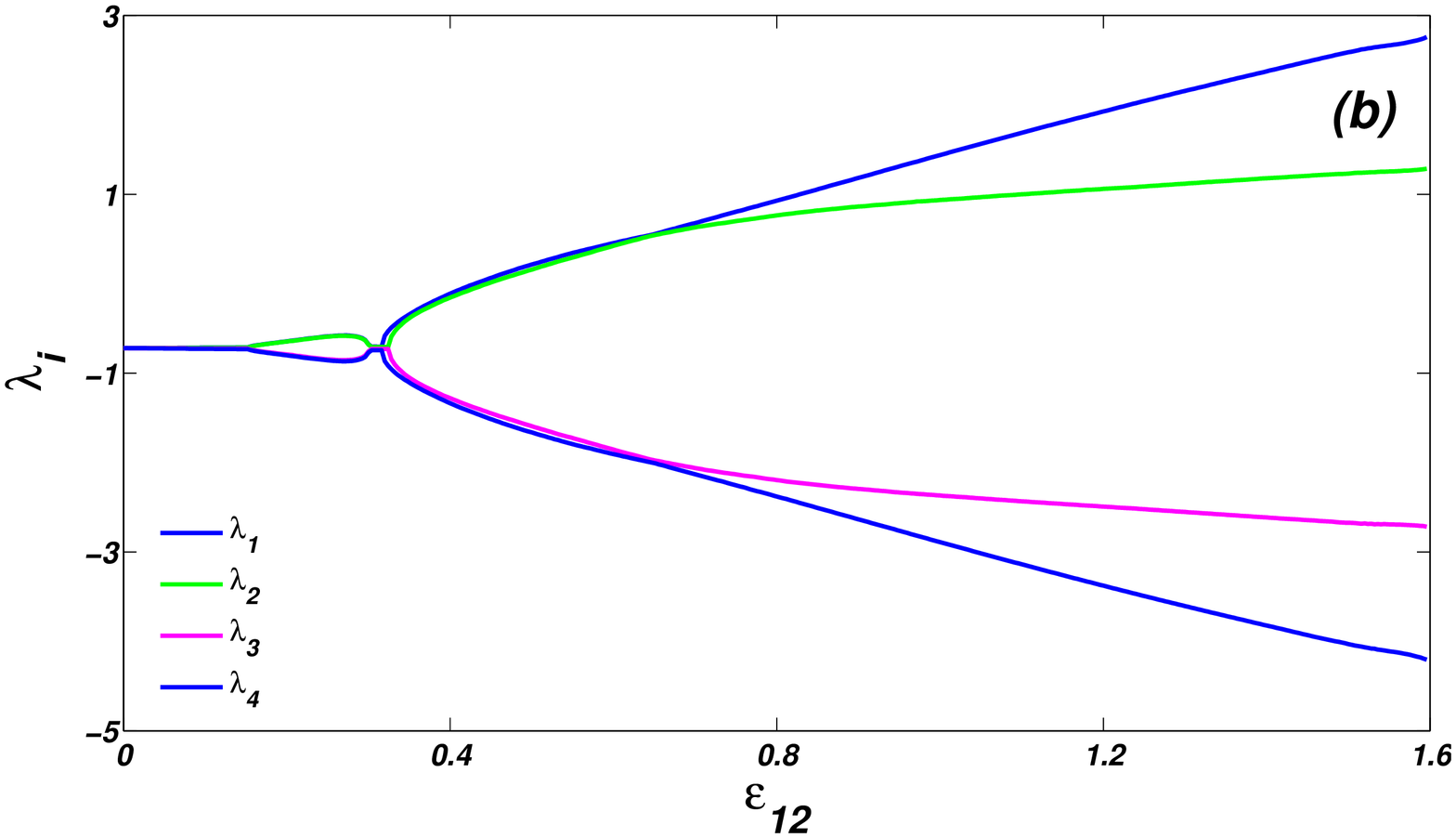}
 \caption{(a) The Lyapunov map of the system (\ref{ker11a})--(\ref{ker11b}) 
in the parametric space ($\epsilon_{12}, \gamma$) where $\gamma=\gamma_{1}=\gamma_{2}$. 
(b) The full spectrum of Lyapunov exponents $\{\lambda_{1}, \lambda_{2}, \lambda_{3}, \lambda_{4}\}$ of the 
system (\ref{ker11a})--(\ref{ker11b}) as a function of the nonlinear coupling parameter $\epsilon_{12}$ and for $\gamma=\gamma_{1}=\gamma_{2}=0.5$. 
The rest of parameters for both figures are:  $\omega_{1}=1$, $\omega_{2}=0.5$,
$\epsilon_{1}=\epsilon_{2}=0.01$, $F_{1}=F_{2}=5$,
 $\Omega_{1p}=3$ and $\Omega_{2p}=2.5$.
The system starts from the initial
conditions: $\mathit{Re}\,a_{1}=10$, $\mathit{Im}\,a_{1}=0$,
 $\mathit{Re}\,a_{2}=10$, $\mathit{Im}\,a_{2}=0$ }\label{genia}
\end{center}
\end{figure}
\begin{figure}
\begin{center}
\includegraphics[width=8cm,height=14cm,angle=0]{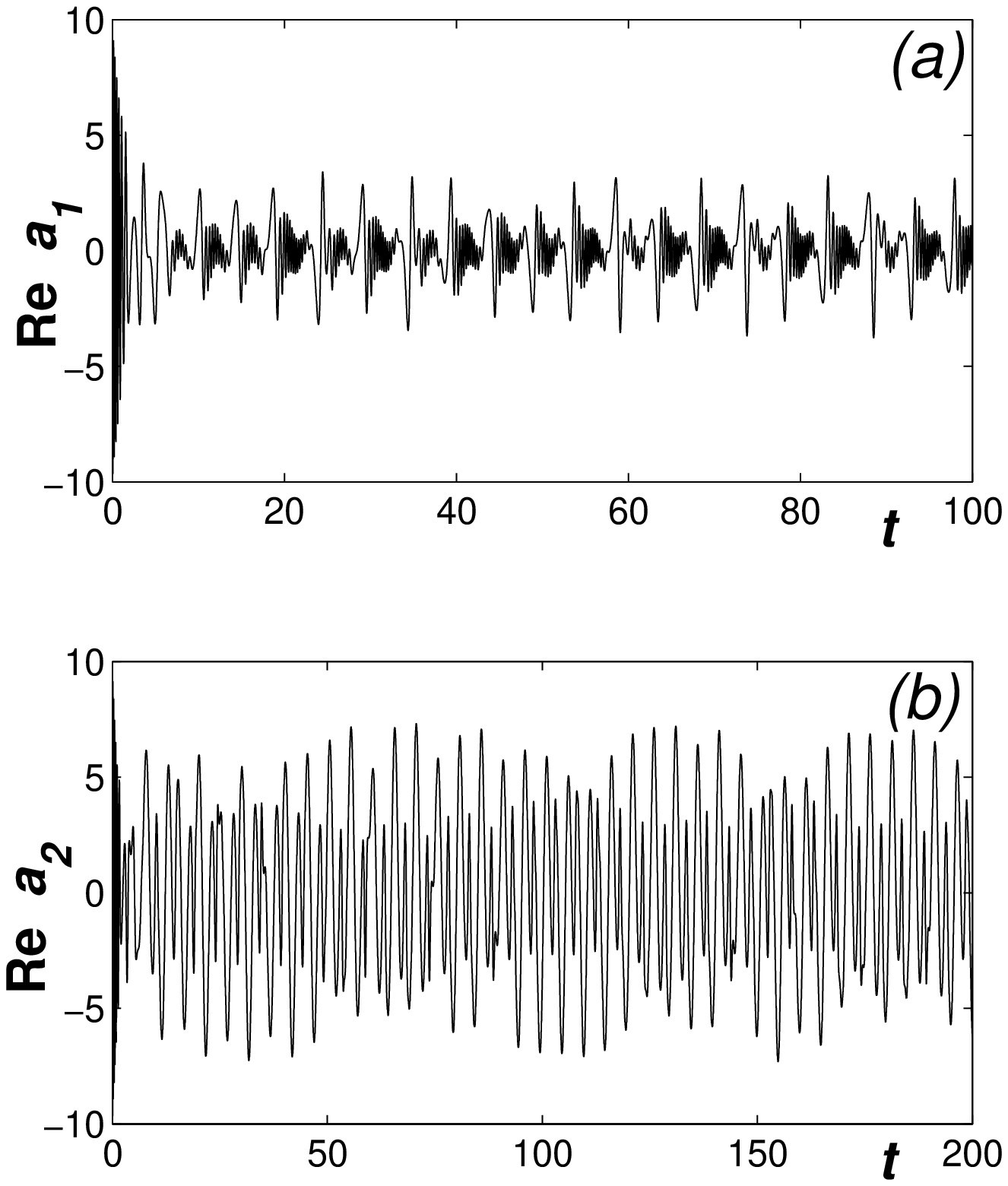}
\caption{Time dependence of $\mathit{Re}\,a_{1}$ (a) and
$\mathit{Re}\,a_{2}$ (b) of the system (\ref{ker11a})--(\ref{ker11b}).
Parameters are the same as in Fig.\ref{genia}(b), and $\epsilon_{12}=0.45$. 
Chaotic beats. }
\label{eps}
\end{center}
\end{figure}
\subsection{Generation of chaotic beats}

Globally, the behaviour of system
(\ref{ker11a})--(\ref{ker11b}) is presented in Fig.\ref{genia}(a) 
showing the Lyapunov map in the parameters space 
($\epsilon_{12}, \gamma$). We find that strong chaotic
behaviour of the system is much common that for the single Kerr system.
We also notice that if we increase the dumping in the system we must
increase the coupling between subsystems to achieve the chaotic behaviour.
The full spectrum of the Lyapunov exponents $\{\lambda_{1},
\lambda_{2}, \lambda_{3}, \lambda_{4}\}$ versus $\epsilon_{12}$
shows the regions of order or chaos in the cross section of the map for the damping parameter $\gamma=\gamma_{1}=\gamma_{2}=0.5$ (Fig.\ref{genia}(b)). 
If $\lambda_{1}>0$ then the system is chaotic, and if $\lambda_{1}\leq 0$, 
it behaves periodically. The system with the parameters of Fig. \ref{genia} and for the coupling constant $\epsilon_{12}>1.6$ manifests extremely unstable behaviour and its solutions are divergent to infinity.
There is also a region of hyperchaotic behaviour of
the system in which two highest Lyapunov exponents are positive
(for $\epsilon_{12}>0.44$).

Taking from Fig.\ref{genia}(a)-(b) the appropriate values of the damping constant $\gamma_{1}=\gamma_{2}=0.5$ and the nonlinear coupling between the Kerr oscillators $\epsilon_{12}=0.45$ we can generate
chaotic beats in the system of two nonlinearly coupled Kerr
oscillators (\ref{ker11a})--(\ref{ker11b}), both being initially in the
periodic state. Such chaotic beats are shown in Fig.\ref{eps}
illustrating the time dependence of $\mathit{Re}\,a_{1}$ and
$\mathit{Re}\,a_{2}$. Because the spectrum of Lyapunov exponents 
of the beats shown in Fig.\ref{eps} is 

$\{0.0708, 0.0205, -1.4552, -1.5215 \}$ 

and contains two positive values we can even call it hyperchaotic beats 
~\cite{ISKGPS}.

\section{Conclusion}

The Kerr effect and the Kerr couplers considered in this
paper have great potential in studies and applications of optical
devices like optical fibres or couplers. From this point of view
the main results of this paper are: 1. Tunnelling properties of
periodic attractors (mutual inter-penetration of attractors and
basins of attraction) lead to the possibility of switching of Kerr
oscillator system between different semi-stable attractors by
changing initial condition (Sec.2.3). 2. As shown in Sec.2.4
the dynamics of the Kerr system can be controlled by
appropriate switching of the parameters of the system or the external
field. Temporary changes in parameters also switch the system
between different periodic states. 3. The system with Kerr
nonlinearity is able to generate chaotic beats (Sec.2.5). The
properties of these beats depend on specific choices of
parameters of the system and the external
field. 4. For two coupled Kerr oscillators it is possible
 to change the stability of attractors
of a given sub-system by changing the initial conditions of the other
sub-system and the dynamics of one sub-system can be controlled by 
changing the initial conditions of the other one (Sec.3.3). 5. Moreover, 
chaotic (or hyperchaotic) beats in the system of two coupled Kerr oscillators can be generated (Sec.3.4).

\end{document}